\documentclass[fleqn,usenatbib]{mnras}
\usepackage{amsmath}
\usepackage{graphicx}
\usepackage{amssymb,txfonts}

\def\sw{{\it Swift}}
\def\fe{{\it Fermi}}

\title[GRB 151006A]
{Surprise in simplicity: an unusual spectral evolution of a single pulse GRB 151006A}
\author[R. Basak et al.]{R. Basak$^{1, 2, 3}$\thanks{E-mail: rupal.basak@gmail.com, rupal@camk.edu.pl}, 
S. Iyyani$^{4}$\thanks{E-mail: shabnam@iucaa.in},
V. Chand$^{5}$, 
T. Chattopadhyay$^{6}$,
D. Bhattacharya$^{4}$,
A. R. Rao$^{5}$,
\newauthor
S.V. Vadawale$^{7}$
\\
$^{1}$ The Oskar Klein Centre for Cosmoparticle Physics, AlbaNova, SE-106 91 Stockholm, Sweden \\
$^{2}$ Department of Physics, KTH Royal Institute of Technology, AlbaNova University Center, SE-106 91 Stockholm, Sweden \\
$^{3}$ Nicolaus Copernicus Astronomical Center, Polish Academy of Sciences, Bartycka 18, 00-716 Warsaw, Poland\\
$^{4}$ Inter University Center for Astronomy \& Astrophysics, Pune, India\\
$^{5}$ Department of Astronomy and Astrophysics, Tata Institute of Fundamental Research, Homi Bhabha Road, Mumbai, India\\
$^{6}$ Dept. of Astronomy \& Astrophysics, Pennsylvania State University, University Park, PA 16802, USA\\
$^{7}$ Physical Research Laboratory, Ahmedabad, India\\
}
\begin{document}

\date{Submitted 2017 February 10}

\pagerange{\pageref{firstpage}--\pageref{lastpage}} \pubyear{2017}

\maketitle

\label{firstpage}

\begin{abstract}
We present a detailed analysis of GRB 151006A, the first GRB detected by Astrosat CZT Imager (CZTI). We study the long term spectral evolution by exploiting the capabilities of \emph{Fermi} and \emph{Swift} satellites at different phases, which is complemented by the polarization measurement with the CZTI. While the light curve of the GRB in different energy bands show a simple pulse profile, the spectrum shows an unusual evolution. The first phase exhibits a hard-to-soft (HTS) evolution until $\sim16-20$\,s, followed by a sudden increase in the spectral peak reaching a few MeV. Such a dramatic change in the spectral evolution in case of a single pulse burst is reported for the first time. This is captured by all models we used namely, Band function, Blackbody+Band and two blackbodies+power law. Interestingly, the \emph{Fermi} Large Area Telescope (LAT) also detects its first photon ($>100$\,MeV) during this time. This new injection of energy may be associated with either the beginning of afterglow phase, or a second hard pulse of the prompt emission itself which, however, is not seen in the otherwise smooth pulse profile. By constructing Bayesian blocks and studying the hardness evolution we find a good evidence for a second hard pulse. The \emph{Swift} data at late epochs ($>T_{90}$ of the GRB) also shows a significant spectral evolution consistent with the early second phase. The CZTI data (100--350\,keV), though having low significance ($1\sigma$), show high values of polarization in the two epochs ($77\%$ to $94\%$), in agreement with our interpretation. 
\end{abstract}

\section{Introduction}
Gamma-ray bursts (GRB) are the most intense cosmological explosions marking the 
formation of the stellar mass compact objects (\citealt{Meszaros_2006}). The central 
engine most probably launches a highly relativistic bipolar jet that releases most of 
the burst energy in a few seconds to a few minutes mostly as gamma-rays, known as the
prompt emission (\citealt{Piran_2004_review}). The surrounding material heated by 
the outflow glows afterwards on a much longer timescale of days to months, known as the afterglow.
As the prompt emission occurs close to the burst site, it carries the most important signature 
of the intrinsic properties of the outflowing plasma and the central engine. 

The non thermal GRB spectral shape is primarily described by 
emission processes like, optically 
thin synchrotron emission (\citealt{Tavani1996,Rees1994,Sari1998}), inverse
Compton scattering (\citealt{Ghisellini1999, Ghisellini_etal_2000}). 
The non-thermal models alone, however, face difficulties in explaining
many spectral features across the GRB catalog, 
such as the hard low energy spectral slopes, collection of the spectral
peaks within a small energy range of a few hundred keV and the high efficiency 
in the production of the observed gamma rays.   
This leads to the inclusion of thermal emission from the photosphere 
(\citealt{Crideretal_1997, Ghirlandaetal_2003, Ryde_2004, Ryde_2005}),
thereby representing the very standard fireball model. 
The observed emission can be entirely from the photosphere wherein 
the broadness is contributed by the continuous or localised subphotospheric 
dissipation of the kinetic energy or Poynting flux 
(\citealt{Beloborodov2010, Beloborodov2013, Giannios2012, Pe'er2005, Rees&Meszaros2005, Iyyanietal_2015}),
or may be due to the geometrical effects of the jet emission and its structure 
(\citealt{Lundman2013,Beloborodov2011, Basak_Rao_2015_090618}).
Or, the photospheric emission can be a sub-dominant component 
(\citealt{Burgess_etal_2014, Axelsson_etal_2012, 
Iyyani_etal_2013, Guiriecetal_2011, Guiriecetal_2013}). All the models have certain
pros and cons and confront issues in justifying the constraints obtained from observations. 
The radiation processes of the prompt 
emission, thus, remains the most debated topic in the GRB field 
(see \citealt{Zhang_2014_review, Kumar&Zhang2015, Pe'er2015} for  recent reviews).

On the other hand, the spectral analysis of the prompt emission itself is challenging because of
the rapid spectral evolution and multiple pulse emission which may have large overlaps.
Several studies conducted for GRBs with single/separable pulses 
show definite evolution of the spectral peak ($E_{\rm p}$), either hard to soft (HTS) 
or intensity tracking (IT) (\citealt{LK_1996, Kanekoetal_2006, Luetal_2012, Basak_Rao_2014_MNRAS}). 
However, for GRBs with overlapping pulses, this picture is complicated which is probably affected 
by pulse overlap (e.g., \citealt{Hakkila_Preece_2014}). 
Also, a GRB detector having an unavoidable wide field of view, is background limited 
and suffers from poor spectral sensitivity. 

The spectral study of the prompt emission has been revolutionized in the past 
two decades particularly after the launch of the \emph{Swift} and the \emph{Fermi} satellites.
The GRB Monitor (GBM) and the Large Area Telescope (LAT) of the \emph{Fermi} provide seven decades of
energy band ranging from 8\,keV to $\sim200$\,GeV, (see \citealt{Meeganetal_2009_fermi_gbm, Atwoodetal_2009_LAT}).
The \emph{Swift} Burst Alert Telescope (BAT) being a coded-mask 
instrument provides a good localization and background estimate in 15--150\,keV, while 
the X-ray Telescope (XRT) covers the lower energy range  of
0.3--10\,keV. The localization accuracy of both together with the fast re-pointing capability 
of the spacecraft facilitates the redshift measurements for the majority of GRBs,
see \cite{Gehrelsetal_2004_swift, Burrowsetal_2005_XRT}. It has been 
suggested that the focusing instruments like the XRT, having a
better sensitivity and spectral resolution than the wide-field detectors, can be used at the late 
epochs to identify any additional spectral component with a good statistical significance
(\citealt{Basak_Rao_2015_090618}). The novel strategy here is to exploit the specific ability
of individual detectors at different phases of a GRB, namely, the wide band \emph{Fermi} data at 
the early phase, complemented by the finest spectral data of the XRT at the late prompt 
and/or early afterglow phase (\citealt{Basak_Rao_2015_090618}) and thereby studying the 
spectral evolution until the late epochs. For even longer GRBs, e.g., the ultra-long 
ones, it is also possible to observe them with highly sensitive detectors of \emph{Chandra} 
and wide band as well as fine spectral resolution detectors of \emph{NuSTAR}
(\citealt{Basak_Rao_2015_130925A}).

Another important addition to these studies would be the polarization
which can be used as a diagnostic tool between different models.
The first such detection was reported for GRB 021206 (linear polarization 
degree, $\Pi = 80 \pm 20 \%$), detected by 
{\it Reuven Ramaty High Energy Solar Spectroscopic Imager (RHESSI)}
(\citealt{Coburn&Boggs2003}, however, see \citealt{Rutledge&Fox2004, Wigger2004}). 
This was followed by e.g.,  {\it INTEGRAL} observation ($100$ keV -- 1 MeV) of 
GRB 041219A and GRB 061122, \cite{Kalemcietal_2007, McGlynnetal_2007, Gotzetal_2009, Gotzetal_2013}, 
{\it IKAROS}/gamma ray burst polarimeter (GAP) observation of
GRB 100826A ($\Pi = 25 \pm 15 \%$ and $31 \pm 21 \% $ in different time intervals), 
GRB 110721A ($\Pi = 84^{+16}_{-28} \%$) and GRB 110301A ($\Pi = 70 \pm 22 \%$), see 
\cite{Yonetokuetal_2011, Yonetoku_etal_2012}.  

High polarization degree is generally owed to the observed non-thermal emission.  
Several theoretical models 
have been proposed such as Compton drag (\citealt{Lazzati_etal_2004}), 
synchrotron emission in ordered magnetic fields (\citealt{Nakar2003, Granot_etal_2003}) 
as well as in random magnetic fields (\citealt{Waxman2003}). On the other hand, 
photospheric emission in general is thought to be unpolarised, however, 
the photospheric emission (blackbody alone) from a structured 
jet, observed off axis is predicted to show a polarization degree up to $\Pi = 40 \%$
(\citealt{Beloborodov2011, Lundmanetal_2014}).  
In case of subphotospheric dissipation models, polarised emission is expected
to be observed in soft X-rays from synchrotron emission when dissipation
occurs close to the photosphere. The MeV peak and spectrum above the
peak is not expected to be polarised as they are formed at high optical depths
(\citealt{Lundman_etal_2016}).
% A point to note is that when the emitting region have rotational 
% symmetry around the line of sight, and  if the detector is unable to 
% spatially resolve different parts of the emission region, 
% the observed overall emission would be unpolarised.  
In order to have a significant polarization detection, 
the crucial factor is to have high photon statistics, 
which has been a concern in most of the polarization measurements. 

\begin{table}\centering
\caption{Observation of GRB 151006A with different detectors. Time is counted since the GBM trigger}%\begin{scriptsize}
% \vspace{0.3in}
\begin{tabular}{ccc}

\hline
Detector & Time bins & Time interval (s) \\
\hline
\hline
GBM      & 1--14      & $-2.0$ -- $88.1$ \\
LAT-LLE  & 1--9       & $-2.0$ -- $27.5$ \\
LAT ($>100$\,MeV) & 8 and 11 & $16-20$, $37-53$ \\
BAT            &  15--18        &   $88.2-253.2$               \\
XRT            &  from bin 12        &   $>58.5$               \\

\hline
\end{tabular}
\label{tobs}
% \end{scriptsize}
\end{table}

% 
 
%  \begin{figure*}\centering
%  {\hspace{1.0in}
%  \includegraphics[width=0.55\textwidth]{lc.pdf} 
%  }
%  \caption{Light curve of GRB 151006A extracted from the \fe~, Astrosat/CZTI and \sw~
%  at different energies (labeled in the respective panels).
%  The time intervals chosen for our analysis are shown by dashed lines. 
%  We also mark the GBM trigger time with dot-dashed red line (at $t=0$\,s)
%  and the first LAT photon detection time with a vertical magenta line 
%  with horizontal dashes (at $t=17.5$\,s). The time intervals for polarization
%  measurements (0--16\,s and 16--33\,s) are marked in the CZTI-veto panel 
%  with dotted brown line.
%  }
%  \label{lc}
%  \end{figure*}
% 
\begin{figure*}\centering
\includegraphics[width=\textwidth]{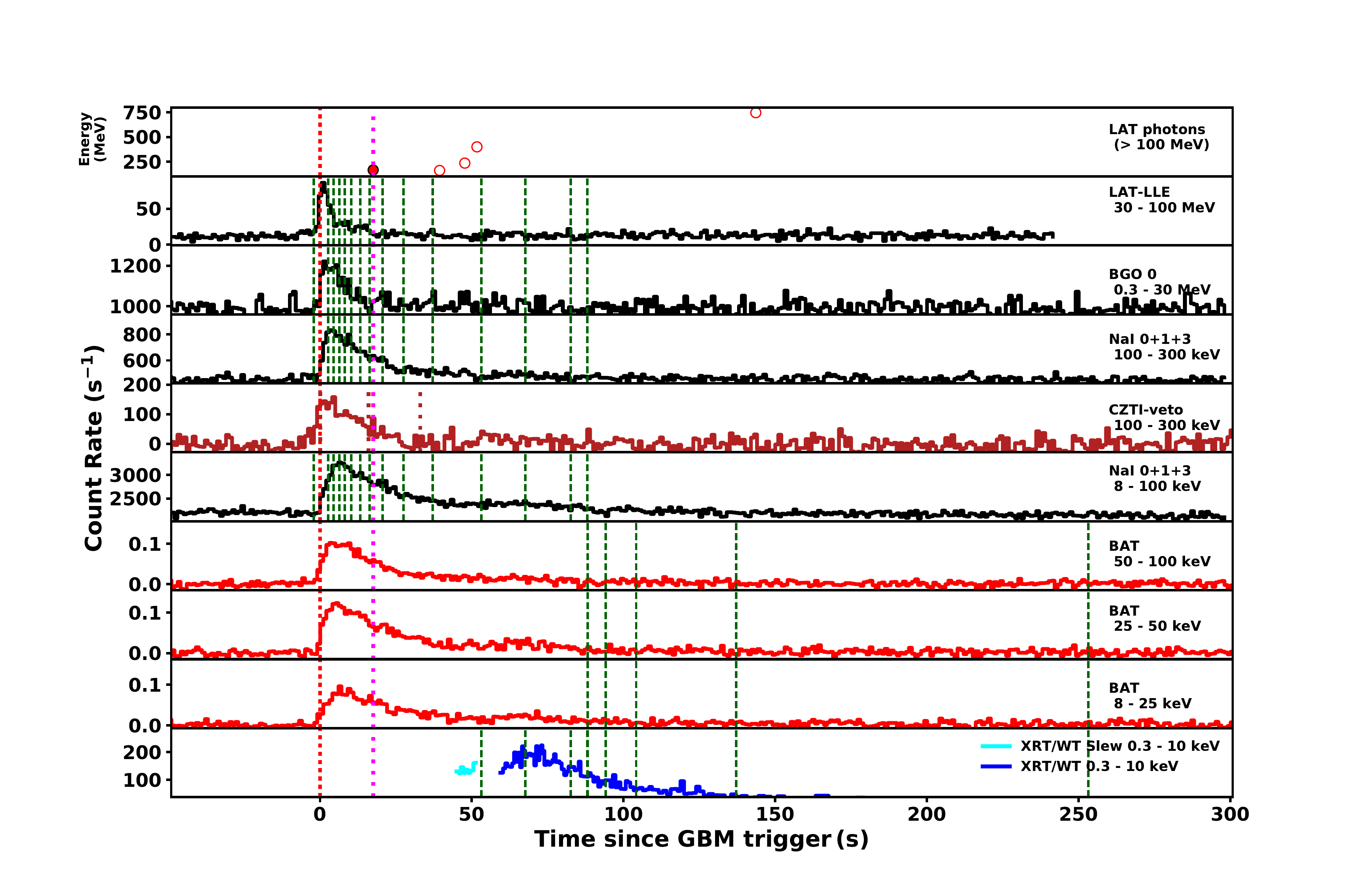} 
\caption{Light curve of GRB 151006A extracted from the \fe~, Astrosat/CZTI and \sw~
 at different energies (labeled in the respective panels).
 The time intervals chosen for our analysis are shown by dashed lines. 
 We also mark the GBM trigger time with dot-dashed red line (at $t=0$\,s)
 and the first LAT photon detection time with a vertical magenta line 
 with horizontal dashes (at $t=17.5$\,s). The time intervals for polarization
 measurements (0--16\,s and 16--33\,s) are marked in the CZTI-veto panel 
 with dotted brown line.}
\label{lc}
\end{figure*}

The CZT Imager (CZTI), on-board the recently launched multi-wavelength Indian mission {\it Astrosat}, 
provides detection in the energy range 20 -- 200\,keV and becomes an open detector above 100\,keV, 
thereby enabling it to detect GRB events. The Veto detector when augmented, raises the detection energy 
level to 600 keV. Thus, when analysed along with the BAT data, the CZTI + Veto will become crucial in 
constraining the spectral peak energies, which is otherwise not generally possible with the BAT 
data alone (\citealt{Raoetal_2016}). The CZTI also possesses X-ray polarization detection capabilities
in the energy range, 100 -- 380\,keV (\citealt{Vadawale_etal_2015}). Thus, with its wide field 
of view, good spectral resolution and polarimetry capability, the CZTI data will be a key addition
to the existing observatories like the {\it Swift} and {\it Fermi}.

In this paper, we present a detailed study of spectral evolution of 
GRB 151006A, the first detected GRB by the CZTI. We follow the strategy of multi-instrument
analysis using the detectors on-board \emph{Fermi} and \emph{Swift} at different 
phases of GRB emission. This is further complemented by 
the polarization data from the CZTI. The paper is organized 
as follows. Section 2 presents the observation by various instruments,
the methodology of the data analysis and the spectral models used; section 3 presents 
the results of the spectral fits 
and the polarization measured by CZTI, finally followed by conclusions and a discussion in Section 4.

\section{Observations and Data Analysis}
On 2015 October 06  at 09:54:57.83 UT, the \fe/GBM (\citealt{151006A_gcn_gbm}) triggered on GRB 151006A.
The burst also triggered many other detectors including the \sw/BAT at 09:55:01 UT (\citealt{151006A_gcn_bat}), 
the \emph{Astrosat}/CZTI (\citealt{151006A_gcn_czti}) at 09:54:57.825 UTC, Konus-Wind at 09:54:57.7 UT 
(\citealt{Konuswind_151006A}) and CALET at 09:54:59.97 UT (\citealt{CALET_151006A}). 

In the current paper, we present the spectral analysis of {\it Fermi} and {\it Swift} data, and the
polarization measurement obtained from CZTI data.
Figure \ref{lc} shows a composite count rate light curve of various detectors on-board the {\it Fermi},
{\it Swift}, and the CZTI of {\it Astrosat} arranged from higher to lower energy bands. These are 
LAT (P8$\_$SOURCE class events, $>100$ MeV), LAT Low energy 
events (LLE, 30 MeV--100 MeV), GBM/BGO detector (300 keV--30 MeV), 
GBM/NaI detectors (100--300\,keV),
CZTI-veto (100--300\,keV),
GBM/NaI detectors (8--100 keV), followed by the
{\it Swift} BAT in 50--100\,keV, 25--50\,keV, 15--25\,keV and 
the {\it Swift} XRT (0.3--10\,keV).
The GBM light curve shows a single pulse having a fast rise and an exponential decay (FRED) 
with a duration ($T_{90}$) of $\sim84$\,s (50--300\,keV) (\citealt{151006A_gcn_gbm}). 
The LLE emission of the burst is coincident with the GBM trigger and peaks simultaneously
with the BGO emission at $ \sim 1.4$ s. The LLE emission is quite significant with a detection 
sigma of 54, and extends for about $\sim 30$ s (\citealt{151006A_LAT_gcn}). The emission consists
of a narrow pulse accompanied by two other smaller pulses towards the end. Thus, LLE emission 
does not show the typical delay that is observed in its onset with respect to the GBM observations 
(\citealt{Ackermannetal_2013_LAT}).  The LAT LLE data bridges the gap between the BGO and LAT 
detections, and thereby helps in constraining high MeV spectral peaks 
(\citealt{Axelsson_etal_2012,Moretti&Axelsson2016}). However, the LAT
(P8R2$\_$source class events, $>100$ MeV) events are observed to arrive 
only $17.5$ s after the GBM trigger. The LAT emission is less significant with
only 5 photons detected in total. The top panel of Fig. \ref{lc} shows the photons
to be associated to the GRB with probability  $ > 0.9$ in red filled circle, 
otherwise in open circles (using {\sc gtsrcprob} of {\it Fermi} science 
tool\footnote{http://fermi.gsfc.nasa.gov/ssc/data/analysis/scitools/overview.html}). 
The delayed onset of the LAT events with respect to GBM emission is consistent
with that observed for long bursts as reported in \cite{Ackermannetal_2013_LAT}.
We note that high energy emission in the energy range, 100 keV -- 30 MeV 
and 30 MeV -- 100 MeV, peaks nearly simultaneously at $\sim 1.4$ s, whereas 
the low energy emission in the energy range, 8 keV - 100 keV, peaks later at $\sim 5.4$ s. 
The BAT light curve also shows a single FRED like pulse with a $T_{90}$ (15 -350 keV) 
of $203.9 \pm 41.6$ s (\citealt{151006A_refined_BAT}). The XRT started observations nearly 
$48.6$ s after the BAT trigger and located an X-ray source at RA = $147.43 \,\rm d$ and DEC = $70.5 \,\rm d$. 
The XRT observed in Window Timing (WT) mode during 55.3--570.8\,s, and in photon counting mode 
afterwards until 114\,ks (both the time counted from the BAT trigger time). The long term XRT light curve 
is best fitted with a double-broken power law with slopes of $-0.5_{-2}^{+0.2}$ until 91\,s, $-2.1\pm0.4$ until
134\,s and $-1.39\pm0.02$ afterwards (XRT repository; \citealt{Evansetal_2009}). In the current analysis we have included  
the XRT data only until $\sim 600\, \rm s$, i.e the part with WT mode observation and nearly coincident 
with the observed BAT emission.\footnote{For a full light curve of the XRT observations, please refer to the 
online repository in the link \url{http://www.swift.ac.uk/xrt_curves/00657750/}}

%{\bf CZTI light curve and detection details.} 
GRB 151006A was also the first GRB detected by the \emph{Astrosat} CZTI, on it's first day of operation
(\citealt{151006A_gcn_czti, Raoetal_2016}). 
Both CZTI and Veto detector light curves also show a single FRED 
like pulse in both the energy ranges 50--200\,keV  and 100 - 500 keV, see Figure 2 in \cite{Raoetal_2016}. 

% \section{}
For the spectral study with the \fe/GBM, we choose three NaI detectors having the highest count rate, namely 
n0, n1 and n3, where the numbering of the NaI detectors follows the usual convention, i.e., `nx' with $x=0-11$. 
As the number of all the NaI detectors are within $x\le5$, we choose the BGO number b0, where `by' 
denotes a BGO detector with $y=0-1$. 
We then use Fermi Burst Analysis GUI v 02-01-00p1 
({\sc gtburst}\footnote{\url {http://fermi.gsfc.nasa.gov/ssc/data/analysis/scitools/gtburst.html}}) 
to extract the spectrum. As the n3 has the 
highest count rate among the chosen NaI detectors, we use it to define the time intervals for time resolved
spectroscopy. We apply a signal-to-noise ratio (S/N) of 20 and find 14 time bins in the interval 
$-2.0-88.1$\,s. The LAT LLE data is also extracted in these time bins until $\sim 27$\,s, following
the standard procedure described in \cite{Ackermannetal_2013_LAT}. 
The LAT P8R2$\_$source class events, $>100$ MeV, events were selected within a $12 \deg$
region centred around the \sw~XRT position. Among the 5 detected LAT photons, only 4  
arrive during the GBM $T_{\rm 90}$ of the burst. Thus, for the temporal analysis, 
depending on the availability of the data, the LAT spectra is extracted in the energy 
range, $100 \,\rm MeV - 1\,\rm GeV$, only for time intervals:  16.3 - 20.5 s (bin 8) and 37 -53 s (bin 11).

The first $88$\,s of the \fe~ data is augmented by the observation with the \sw/BAT ($88-253$\,s) and \sw/XRT 
at later times ($> 58.5$\,s). 
The data of the BAT was extracted following the standard procedure. The data is calibrated using the task
{\sc bateconvert}, followed by constructing a detector plane image using {\sc batbinevt}. The known
bad detectors and the noisy pixels are eliminated by {\sc batdetmask} and {\sc bathotpix}. It is then mask
weighted by {\sc batmaskwtevt}. Finally, the spectrum is extracted with {\sc batbinevt} in a specified 
time interval. The spectrum is corrected by  ray-tracing with {\sc batupdatephakw} and the response 
matrix is generated using {\sc batdrmgen}. The XRT data was extracted using the standard tools provided 
by the UK Swift Science Data Centre (\citealt{Evansetal_2009})\footnote{\url {http://www.swift.ac.uk/burst_analyser/}}.
We extract the spectrum from the WT data with a pileup and exposure map correction. 

As for the spectral model, we first choose Band function (\citealt{Bandetal_1993}), which 
is a broken power law with two photon indices, $\alpha$ and $\beta$ and a peak, $E_{\rm p}$ 
in the $\nu F_\nu$ representation. A majority of GRB 
data is consistent with this function (e.g., \citealt{Gruberetal_2014, Goldstein_etal_2013}).
As we will show in Section~\ref{spectral_evolution}, the evolution of the $E_{\rm p}$ of the 
Band function with time shows a sudden jump. As GRB spectrum has been found to have multiple spectral components, this 
sudden jump can as well represent another peak in the spectrum which is not captured 
by the single peak Band function. In order to check that the spectral variation is real, we then 
use two models, Band function + Blackbody (Band + BB), e.g., 
\cite{Guiriecetal_2011, Axelsson_etal_2012, Guiriecetal_2013} and 
Two blackbodies + power law (2BB + PL), e.g., \cite{Basak_Rao_2014_MNRAS,
Basak_Rao_2015_090618, Iyyanietal_2015}. In addition, at the later phase
where the data does not allow to put a constrain on the high energy power law,
we use instead a blackbody + power law (BB + PL) model.

The spectral analysis is carried out in {\sc xspec} version: 12.9.0. For the analysis involving \fe/GBM 
and LAT data, PG-Statistic is used (\citealt{Greiner_etal_2016}) and that involving \sw/BAT and XRT data, 
$\chi^2$ statistic\footnote{\url {http://swift.gsfc.nasa.gov/analysis/bat_swguide_v6_3.pdf}} is used.
All the errors on the fit parameters are quoted at $1\sigma$ (nominal 68\% confidence).

\begin{table*}\centering
\caption{Parameters of time resolved spectral fitting with Band model. Bin 8 and 11 have a 
simultaneous coverage with the LAT data marked $^{(a)}$}%\begin{scriptsize}
% \vspace{0.3in}
\begin{tabular}{ccccccccc}

\hline
Bin \# & Time interval (s) & $\alpha$ &  $\beta$ & $E_{\rm peak}$\,(keV)  & $N_{\rm Band}\,(10^{-3})$ &  PG-Stat (dof)  \\ 
\hline
\hline 
$1$        &$-2.0$ -- $ 2.7$ & $   -0.90_{ -0.06}^{+ 0.05}$    & $   -2.7_{-0.2}^{+0.1}$   & $ 3082_{  -631}^{+ 1110}$    & $    3.4_{-0.2}^{+ 0.2}$  &  $  489.8\,(477)$   \\
$2$        &$ 2.7$ -- $ 4.5$ & $    -1.13^{+0.04}_{-0.05}$     & $   -2.4^{ +0.1}_{- 0.2}$ & $ 2325^{+3475}_{-620}$       & $ 10.9_{- 0.5}^{ +0.4}$   &  $ 457.2\,(477)$   \\
$3$        &$ 4.5$ -- $ 6.3$ & $  -1.10^{+0.11}_{-0.04}$       & $   -2.5^{ +0.2}_{- 0.2}$ & $ 921^{ +9036}_{- 416}$      & $ 12.3^{ +2.0}_{- 0.6}$   &  $  532.3\,(477)$ \\
$4$        &$ 6.3$ -- $ 8.2$ & $  -1.14^{+ 0.29}_{- 0.09}$     & $   -2.3^{+ 0.2}_{- 0.2}$ & $ 625^{+670}_{- 424}$        & $ 13.3^{+ 7.7}_{- 1.5}$   &  $ 499.2\,(477)$ \\
$5$        &$ 8.2$ -- $10.3$ & $  -1.25^{+ 0.10}_{-0.03}$      & $   -2.5^{+ 0.1}_{- 0.2}$ & $ 1292^{+ 669}_{- 596}$      & $ 10.5^{+ 1.4}_{- 0.4}$   &  $ 507.8\,(477)$ \\
$6$        &$10.3$ -- $13.2$ & $ -1.18^{+0.09}_{- 0.08}$       & $   -2.4^{ +0.1}_{- 0.2}$ & $ 582^{ +370}_{- 191}$       & $ 9.9^{+ 1.3}_{-1.1}$     & $   471.2\,(477)$\\
$7$        &$13.2$ -- $16.3$ & $  -1.16^{+ 0.11}_{-0.09}$      & $   -2.3^{+ 0.1}_{-0.1}$  & $ 411^{ +237}_{- 135}$       & $ 10.1^{ + 1.8}_{- 1.3}$  & $  529.5\,(477)$\\ 
$8^{(a)}$  &$16.3$ -- $20.6$ & $  -1.21^{+ 0.13}_{-0.12}$      & $   -2.5^{+ 0.1}_{-0.2}$  & $ 492^{+ 619}_{-204}$        & $ 7.6^{+ 1.7}_{-1.3}$     & $  574.9\,(487)$\\
$9$        &$20.6$ -- $27.5$ & $  -1.08^{+0.22}_{- 0.30}$      & $   -2.2^{+ 0.1}_{-0.3}$  & $ 208^{+950}_{-86}$          & $7.3^{+3.3}_{-2.9}$       & $   612.8\,(477)$\\
$10$       &$27.5$ -- $37.1$ & $   -1.34^{+ 0.16}_{-0.10}$     & $   -1.6^{+0.1}_{-0.1}$   & $ 580^{+6619}_{- 397}$       & $3.1^{+0.8}_{-0.4}$       & $   640.5\,(469)$\\
$11^{(a)}$ &$37.1$ -- $53.2$ & $   -1.34^{+ 0.06}_{- 0.06}$    & $   -3.1^{+0.4}_{- 0.3}$  & $ 8318^{+7462}_{-4801}$      & $1.6^{+0.1}_{-0.1}$       & $   747.4\,(479)$\\ 

\hline
\end{tabular}

\label{t_band}
% \end{scriptsize}
\end{table*}

\subsection{Effective area correction:} 
The different detectors used for the spectral analysis are expected to have some differences
in the calibration. We multiply a constant factor for each detector used for a spectral fitting
for the effective area correction (EAC). However, as each time-resolved data has a limited number 
of counts, the EAC constant may not be well constrained. Hence, we adopt the following procedure.

First, note that it is sufficient to determine the correction factor of all the detectors relative
to one of the detectors. Hence, we freeze the EAC constant value corresponding to the detector having 
the highest count rate to 1, while that of the other detectors are made free. We then load the time-resolved 
data of all the detectors in all the time bins simultaneously. Then the EAC constant parameter of 
the individual detectors are linked throughout all the time bins. For example, let us assume that 
we have three detectors and four time bins, with detector 1 having the highest count rate. 
We load $3\times4=12$ data groups simultaneously. Group 1--3 correspond to time bin 1, group 4--6 
correspond to time bin 2 and so on. Now, the constant parameter of group 1 is fixed to 1, while that of 
group 2 and 3 are free. From the next time bin onward this parameter is linked to the 
corresponding value of time bin 1. Thus, the constant parameter of the group 4 is 
set equal to that of the group 1, that of the group 5 is set equal to that of the group 2,
and so on. Note that this procedure is allowed as the EAC of the detectors cannot vary with time.
On the other hand, linking this parameter throughout all the time bins enable us to determine 
the value with much higher accuracy than that obtained by individual fits.

In the present study, as the n3 detector has the highest count rate, we freeze the EAC constant 
of this to 1. As the LAT-LLE data is available until bin 9, we use the procedure until this time 
bin. For the various models we use, we find that the EAC constant of all the detectors of GBM give quite 
similar values. These are --- n3: 1 (fixed), n0: $0.95\pm0.03$, n1: $0.93\pm0.03$, b0: $1.0\pm0.1$.
The EAC constant corresponding to the LAT-LLE was found to widely differ between models. The value 
is sometimes unrealistically low ($<0.3$). Based on the current understanding of the different 
detectors of the \fe,~ the EAC constant factor is not expected to differ by more than $30\%$.
We find that the Band + BB gives the most realistic value of this factor 
in the range $0.40-1.02$. Note that due to much less number of spectral bins, the 
constant factor is not well constrained even after using the above mentioned procedure. As the value is 
consistent with having no correction, we freeze the EAC constant factor corresponding to the LAT-LLE 
and the LAT data ($>100$\,MeV) 
to 1 for all spectral fitting. For the XRT analysis, we find the constant factor in 
the range 0.8--1.2, and hence, we freeze it to 1. We do not apply any cross-calibration between the XRT 
and the BAT.

\section{Results}
\label{results}

\subsection{Spectral evolution during the prompt emission}
\label{spectral_evolution}

\begin{figure}\centering
{
\includegraphics[width=\columnwidth]{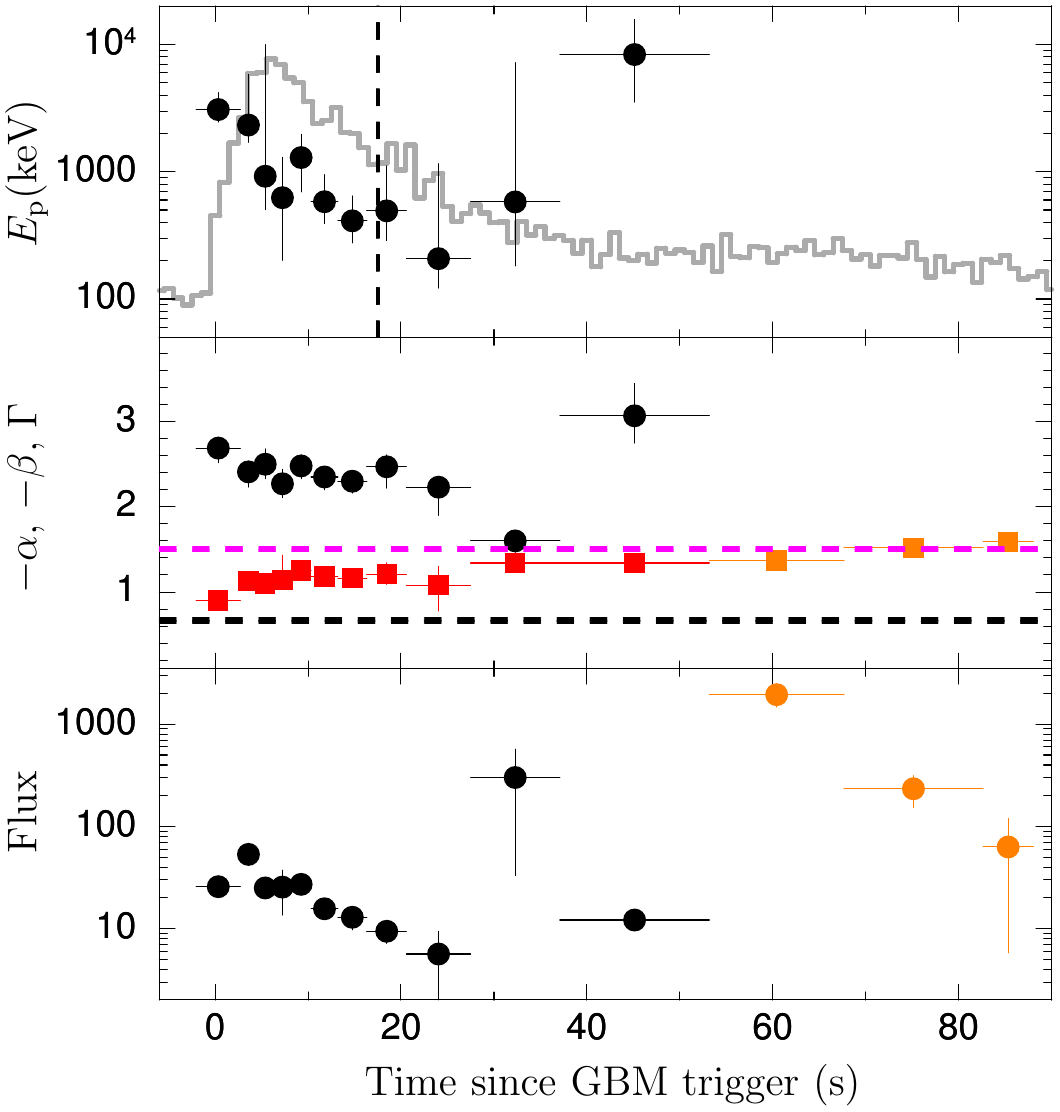} 
}
\caption{Evolution of the parameters of the Band model fitted to the time-resolved \emph{Fermi}/GBM,
LAT and XRT data. For Bin 12--14, we show the parameters of the power law model fits in orange 
symbols. Top to bottom --- Panel 1: $E_{\rm p}$; Panel 2: Photon index $\alpha$ 
(red filled boxes), $\beta$ (black filled circles) of the Band function, and the 
power law index, $\Gamma$ (orange filled boxes); 
Panel 3: Energy flux in units of $10^{-7}$\,erg\,cm$^{-2}$\,s$^{-1}$. The light curve in energy range,
8 keV -- 100 keV, is shown in gray in Panel 1.
The detection of the first LAT photon at 17.5\,s is marked by a vertical dashed line in Panel 1.
In Panel 2, the synchrotron fast cooling photon index of 3/2 and the slow cooling photon index of 2/3 are marked by 
horizontal dashed lines.
}
\label{band_parm}
\end{figure}

\begin{figure}\centering
{
\includegraphics[width=\columnwidth]{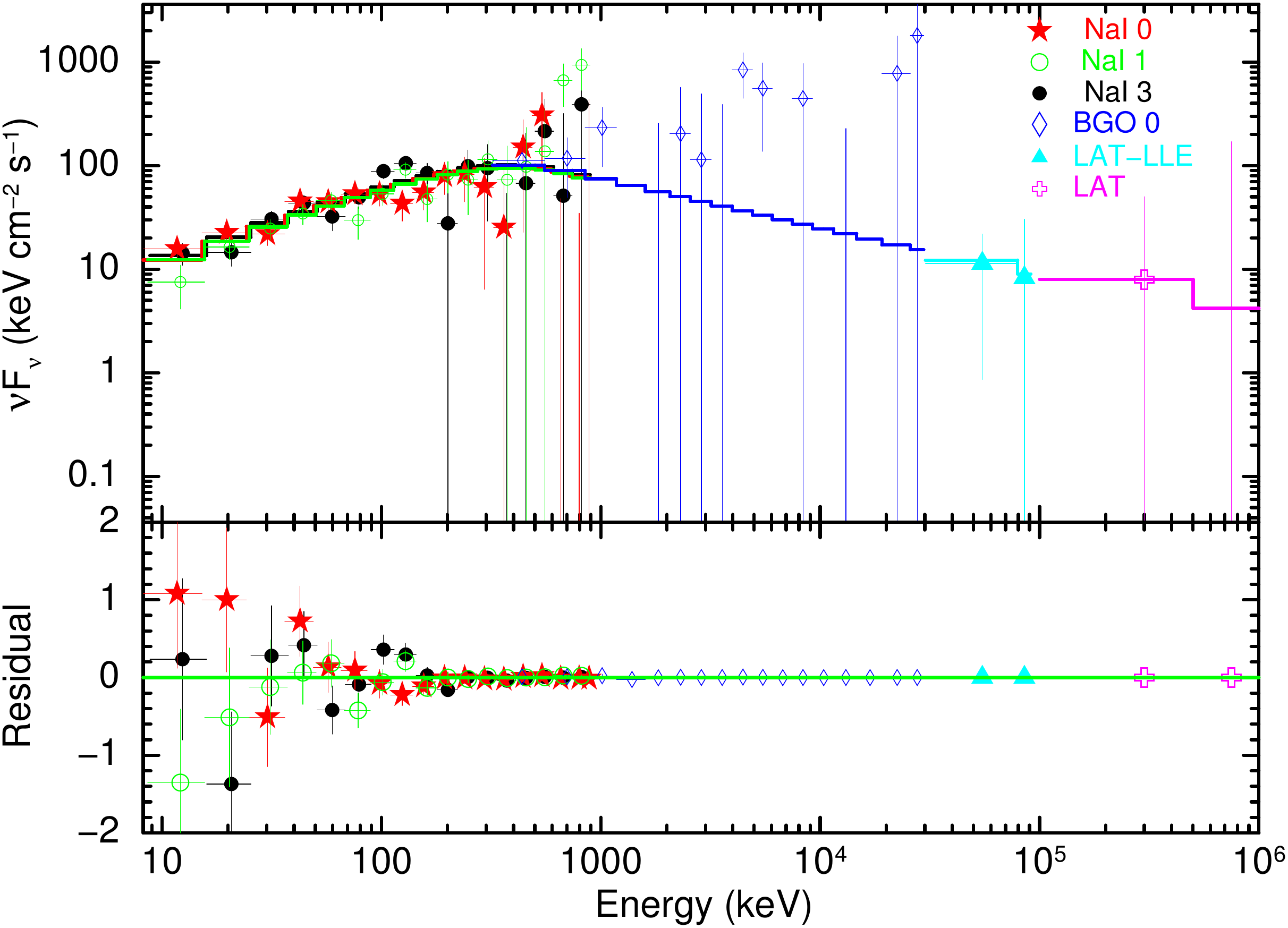} 
}
\caption{Wide band spectrum of Bin 8 fitted with the Band function. Markers
(and colours) used for different detectors are shown in the legend.
}
\label{wide_spectrum}
\end{figure}

\begin{figure}\centering
{
\includegraphics[width=\columnwidth]{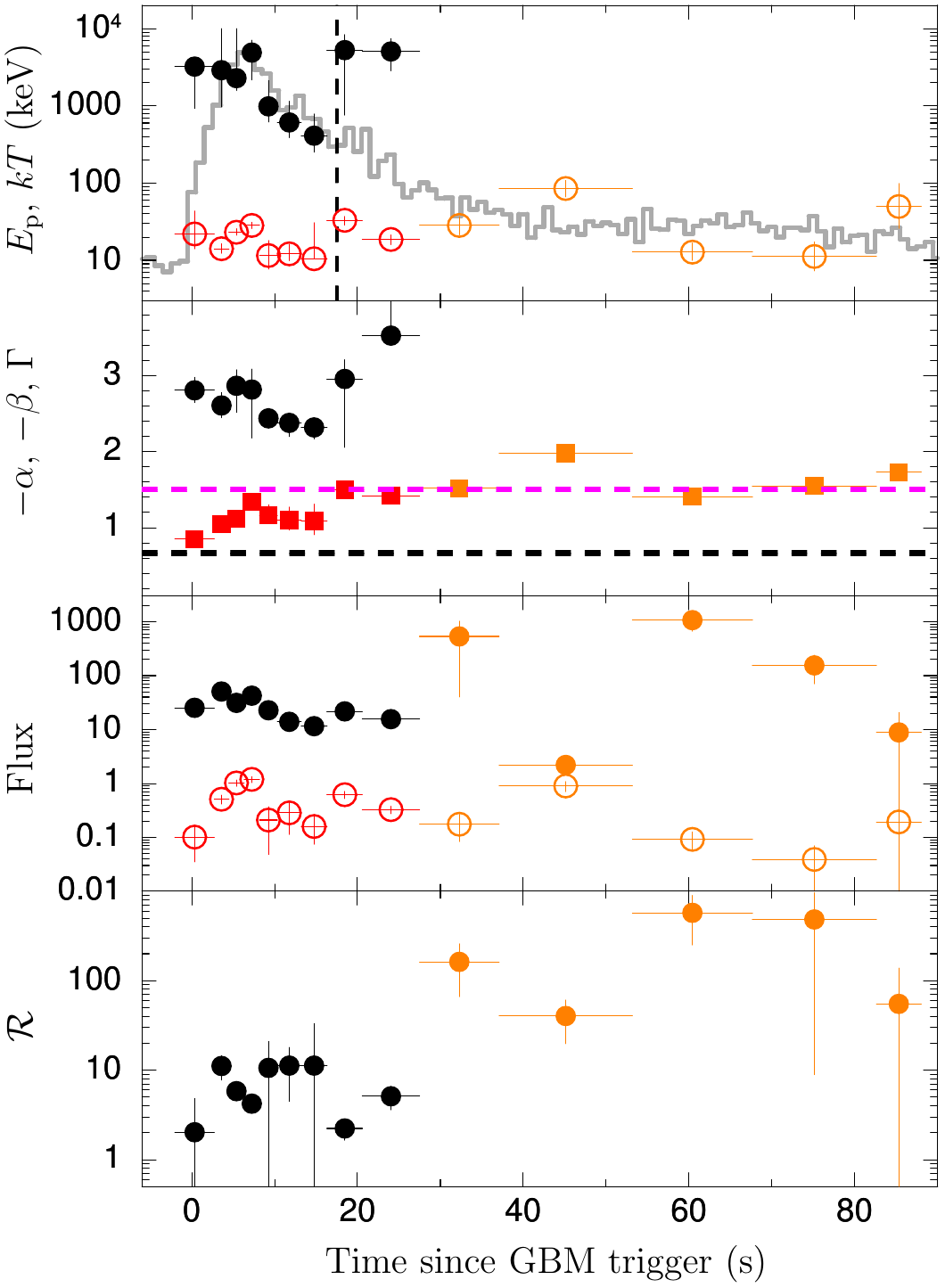} 
}
\caption{Evolution of the parameters of the Band + BB model fitted to the time-resolved \emph{Fermi}/GBM,
LAT and XRT data. For Bin 10 --14, the parameters of the BB + PL model fits are shown in orange 
symbols. From top to bottom --- Panel 1: $E_{\rm p}$ of the Band (black filled circles) and $kT$ 
of the blackbody (red open circles); Panel 2: Photon index $-\alpha$ (red filled boxes), 
$-\beta$ (black filled circles) of the Band function, and the power law index, $\Gamma$ (orange filled boxes);
Panel 3: Energy flux in units of $10^{-7}$\,erg\,cm$^{-2}$\,s$^{-1}$
for Band (black filled circles), BB (red open circles) and power law (orange filled circles);
Panel 4: $\cal{R}$ in units of $10^{-21}$. The light curve in energy range, 8 keV - 100 keV, is shown in grey in Panel 1.
The detection of the first LAT photon at 17.5\,s is marked by a vertical dashed line in Panel 1.
In Panel 2, the synchrotron fast cooling photon index of 3/2 and the slow cooling photon index of 2/3 are marked by 
horizontal dashed lines.}
\label{babb_parm}
\end{figure}

\begin{figure}\centering
{
\includegraphics[width=\columnwidth]{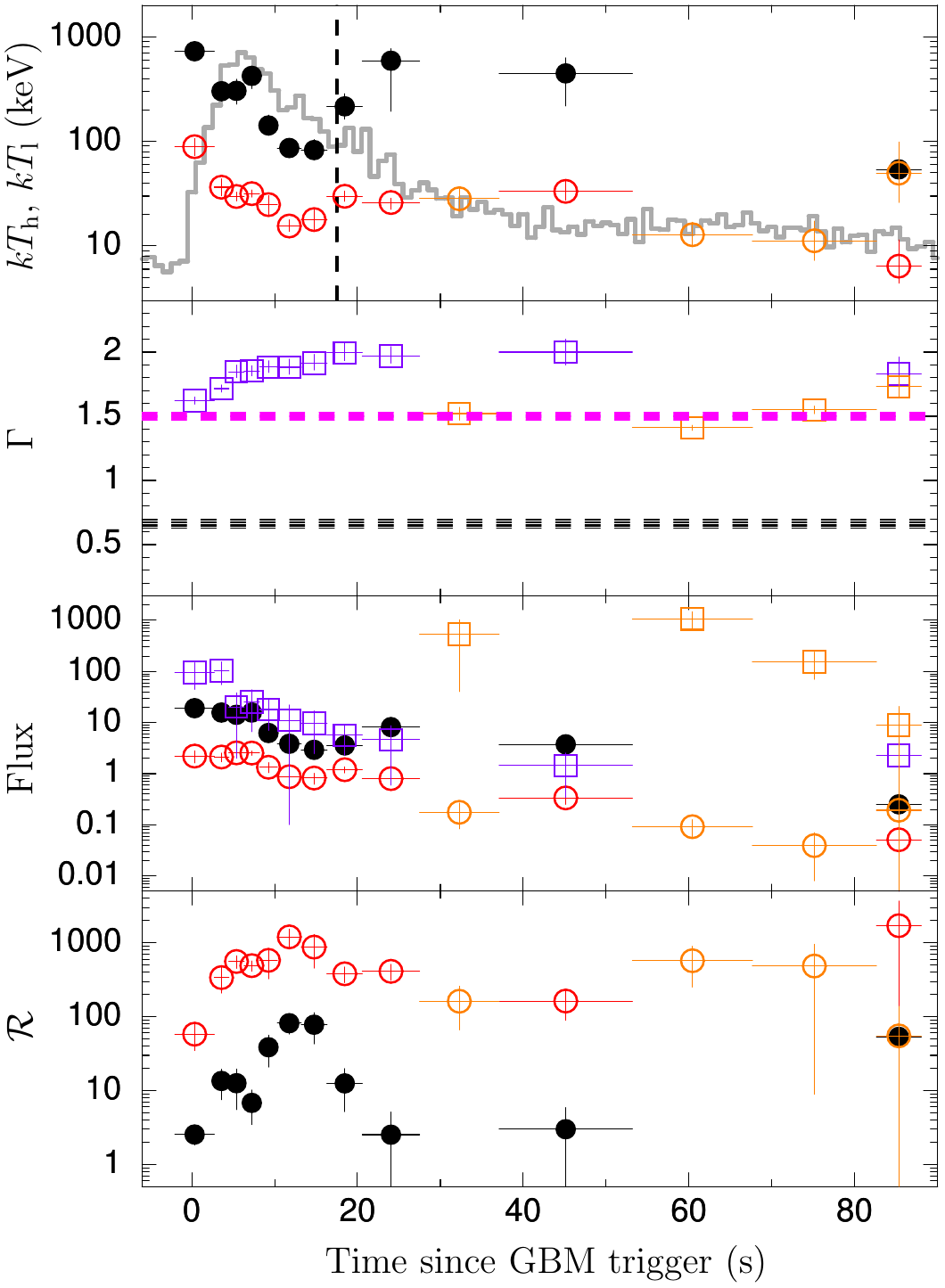} 
}
\caption{Evolution of the parameters of the 2BB + PL and BB + PL model fitted to the time-resolved \emph{Fermi}/GBM,
LAT and XRT data are shown. The parameters of the high energy blackbody, low energy blackbody 
and the power law component of the 2BB + PL model
are shown in black filled circles, red open circles and violet open boxes respectively.
The parameters of  BB + PL model fits are shown in symbols of orange colour.
From top to bottom --- Panel 1: Temperature of the blackbodies; Panel 2: Power law photon index, $\Gamma$; 
Panel 3: Energy flux of each component in units of $10^{-7}$\,erg\,cm$^{-2}$\,s$^{-1}$;
Panel 4: $\cal{R}$ in units of $10^{-21}$. The light curve in energy range, 8 keV - 100 keV, is shown in grey in Panel 1. 
The detection of the first LAT photon at 17.5\,s is marked by a vertical dashed line in Panel 1.
%The vertical dashed line in panel 1, signifies the time ($\sim 18 \, \rm s$) beyond which a deviation in the HTS evolution of spectral peak is observed. 
}
\label{2bbpl_parm}
\end{figure}

The results of the Band fits are shown in Table~\ref{t_band} and Fig.~\ref{band_parm}.
An example of Band fit is shown for Bin 8, where the wide band data including that of 
the LAT is available, see Fig.~\ref{wide_spectrum}.
The parameters could be constrained until Bin 11, and afterwards the Band turns to be a 
single power law. Note that the $E_{\rm p}$ evolution in Panel 1 of Fig.~\ref{band_parm} shows a 
hard-to-soft (HTS) evolution until Bin 9 and then the value starts to increase and reaches a few MeV. 
When the spectral evolution is compared with the 8 keV --100\,keV light curve, shown in the background, the
evolution does not seem to show an intensity-tracking (IT) trend either. Note from Fig.~\ref{lc} that none 
of the light curves in any energy band show any new pulse at this phase. 
However, interestingly, we note that the first LAT photon ($ > 100\, \rm MeV$) arrives at 17.5\,s,
which is marked by a dashed line in both Fig.~\ref{lc} and Fig.~\ref{band_parm}. 
The observed change in the spectral peak of the Band function
takes place during this time. This thereby 
suggests either an onset of a second hard pulse related to the prompt emission 
or the beginning of the afterglow phase (see Section~\ref{second_pulse} and 
Section~\ref{disc} for a detailed discussion).

In the second panel of Fig.~\ref{band_parm}, we show the evolution of the photon indices of the Band 
function as well as that of the power law. Negative values of $\alpha$ and $\beta$
are shown in order to match the convention of the photon index $\Gamma$ of the power law model i.e., 
$N(E)\propto E^{-\Gamma}$. 
The $\alpha$ values are found to be softer than the line of death of synchrotron emission (\citealt{Preeceetal_1998})
i.e $\alpha = -0.67$, corresponding to slow cooling synchrotron emission,
throughout the burst duration. In accordance to what is typically observed 
(\citealt{Kanekoetal_2006}), $\alpha$ is found to get softer with time, tending to values, 
$\alpha = -1.5$, consistent with fast cooling synchrotron emission (marked as pink horizontal dash
line in Fig.~\ref{band_parm} ).  The value of $\beta$ also decreases gradually over time. 
As a result of which, at later times, it becomes difficult to constrain the spectral peak and 
spectrum is then modelled by a simple power law. The power law index, $\Gamma$ has values equal
to $1.5$. Such hard values of $\Gamma < 2$ suggest either a spectral peak or cutoff to lie beyond 
the observed energy window, e.g \cite{Gonzalezetal_2003}.

In the third panel we show the evolution of bolometric energy flux in 0.1\,keV -- 100\,GeV.
We however note that the bolometric 
value inherently assumes that the same model can be extended in both lower and higher energies. While an observed 
flux would not have such assumption, it can underestimate the powerlaw flux. Also, as we have used different 
detectors at different phases of the GRB, the bolometric flux automatically provides a uniform band for all
observation. The flux at any desired energy band can be easily calculated using the parameters given in the tables.
The Band flux smoothly decreases until bin 9.
Note that the power law model as well as in case of the Band function (bin10), where the Band nearly tends 
to be a power law ($\beta$ is nearly equal to $\alpha$), show an increased flux in comparison to the 
otherwise smooth evolution. This is due to the fact that in other bins, the Band function has a 
steeper slope at higher energies than the power law which does not have a break. 
Consequently, in these bins the fluxes are over estimated.

The apparent jump in $E_{\rm p}$ evolution as found above calls for a detailed study. This is solely a spectral 
variation as the underlying lightcurve does not show any change. It is known that GRB spectrum can 
have multiple components with two peaks e.g., Band + BB, or 2BB + PL. If one of the 
components is dominant, it is possible that our single component Band function picks up that one
leading to an artificial $E_{\rm p}$ variation. Hence, in order to check that the variation is 
physical we use these two models. The corresponding
spectral evolutions are shown in Fig.~\ref{babb_parm} for Band + BB and Fig.~\ref{2bbpl_parm}
for 2BB + PL model. Whenever the curvature at high energies cannot be constrained,
we use the simpler BB + PL model, which these two models would converge to. 
For the parameters of all the fits, see Appendix~\ref{table_spectral}.
For bins 12--14, we include simultaneous XRT observation as well. 
As soft X-rays suffer from absorption at the source and the Galaxy, we include two absorption 
terms ({\sc TBabs} in {\sc Xspec}) in the model. The equivalent hydrogen column density ($n_{\rm H}$) of the 
Galactic term is fixed to $7.74\times10^{20}$\,atoms cm$^{-2}$. Then we link the $n_{\rm H}$ of the 
source absorption term between the time bins. We obtain $n_{\rm H}=(5.1\pm0.6)\times10^{21}$\,atoms cm$^{-2}$.
For subsequent fitting with the XRT data, we freeze the source  $n_{\rm H}$ at this value as this cannot 
evolve with time and the linking of the parameter gives good confidence on the obtained value.

From the upper panels of Fig.~\ref{babb_parm} and Fig.~\ref{2bbpl_parm} we see a very similar 
variation of the spectral peak as before. This indeed shows that the sudden jump in the peak 
energy is not an artefact of the adopted model. 
The next two panels of the Figures are same as 
Fig.~\ref{band_parm}. In the fourth panel, we also show the 
evolution of the parameter $\cal{R}$$\equiv(F_{\rm BB}/\sigma T^4)^{1/2}$ of the blackbody component. 
${\cal{R}}$ represents the effective transverse size of the emitting region (i.e photosphere) 
provided the bulk Lorentz factor of the outflow, $\Gamma_{\rm B} \gg 1/\theta_{\rm j}$ where
$\theta_{\rm j}$ is the jet opening angle (\citealt{Ryde_Pe'er_2009}).
For Band + BB note that $\cal{R}$ does not show a linear increase in contrast to what is typically observed
(\citealt{Ryde_Pe'er_2009,Ryde_2004,Ryde_2005}), instead only exhibits an overall increment throughout.
On the other hand, the variation of $\cal{R}$ for both the thermal components of 2BB + PL 
model exhibit a similar jump corresponding to the evolution of $E_{\rm p}$.

% \subsubsection{Band function + Blackbody:}

\begin{figure}\centering
{
\includegraphics[width=\columnwidth]{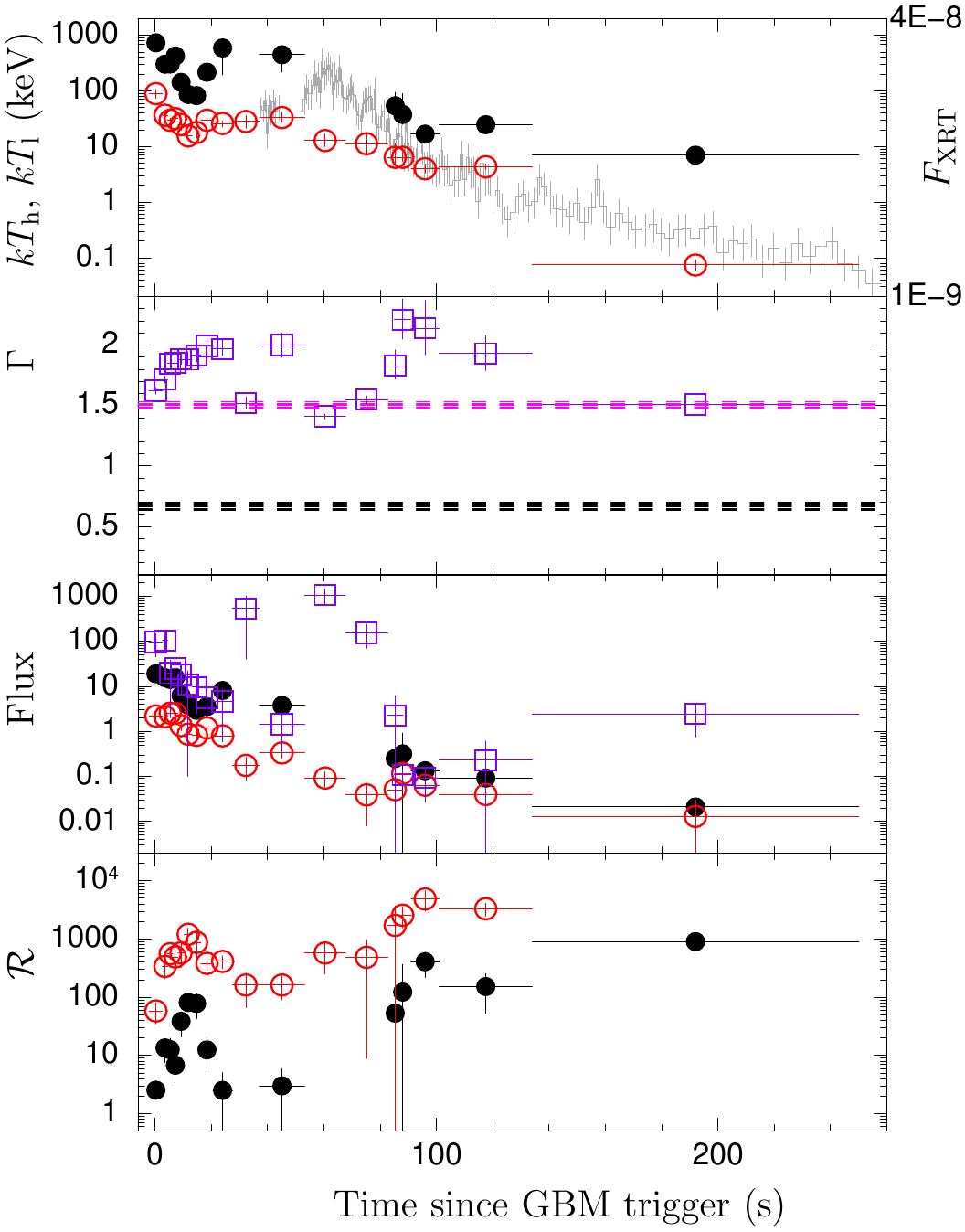} 
}
\caption{Long time evolution of the parameters of the 2BB + PL model are shown. 
The symbols used are the same as in Fig.~\ref{2bbpl_parm}. The data after 88\,s 
correspond to Table~\ref{t_late_2bbpl}. The 0.3--10 keV XRT flux
(erg\,cm$^{-2}$\,s$^{-1}$) is shown in the background of Panel 1
and the corresponding vertical scale is shown on the right.
The value of $\cal{R}$ of the lower-temperature blackbody is not shown for 
the last time interval, as due to a very low temperature, we obtained an 
unrealistically high value.
}
\label{2bbpl_late_evolution}
\end{figure}

\begin{figure}\centering
{
\includegraphics[width=\columnwidth]{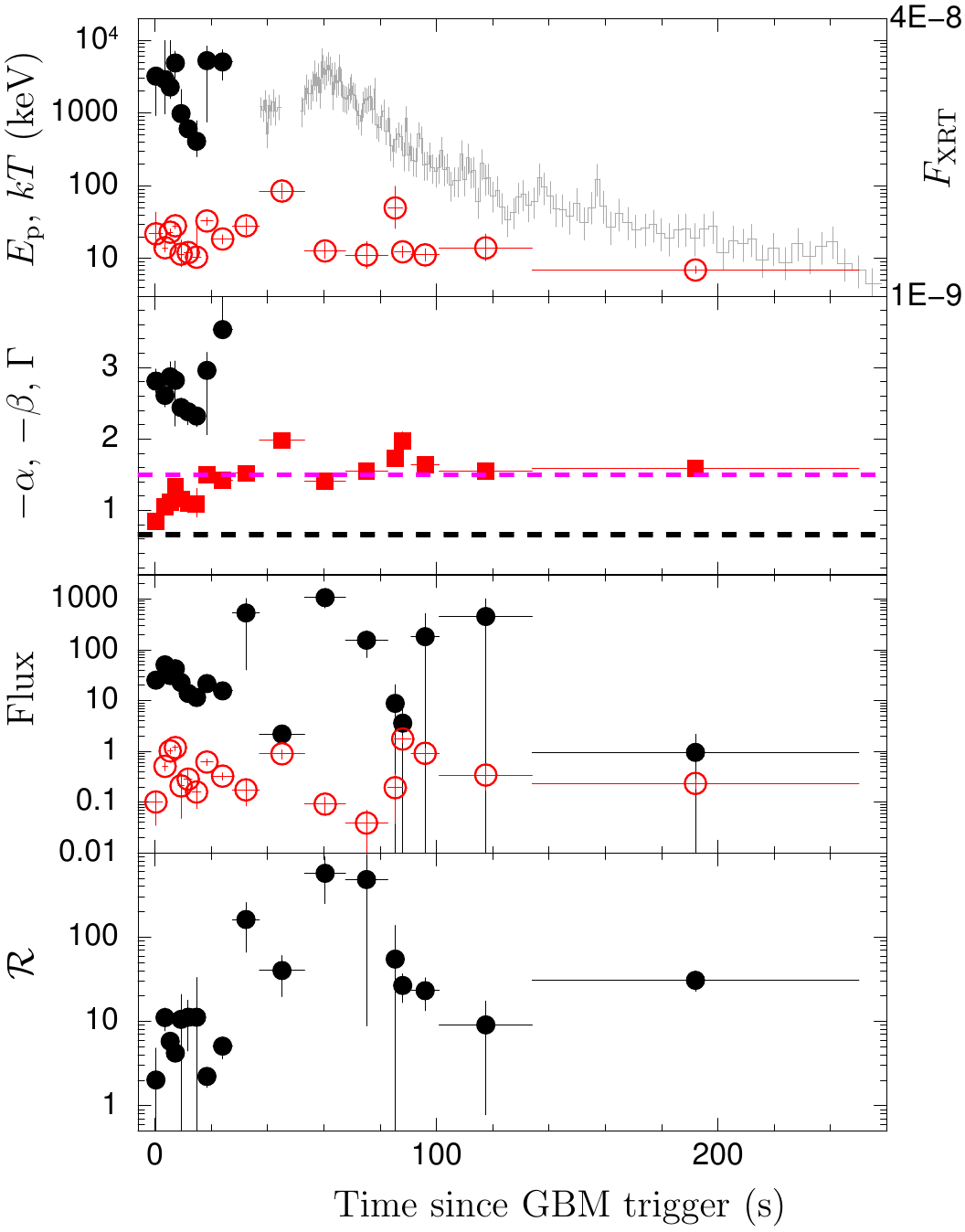} 
}
\caption{Long time evolution of the parameters of the Band+BB model (Bin 1--9) and 
BB+PL model (Bin 10--18) are shown.
The symbols used are the same as in Fig.~\ref{babb_parm}. 
The 0.3--10 keV XRT flux
(erg\,cm$^{-2}$\,s$^{-1}$) is shown in the background of Panel 1
and the corresponding vertical scale is shown on the right.
}
\label{babb_late_evolution}
\end{figure}

\subsection{Spectral evolution at late times ($> 88 \,\rm s$)}
\label{late_phase}

\begin{table*}\centering
\caption{Parameters of time resolved spectral fitting of the joint BAT and XRT data at late times.
We also show the $\chi^2$ of the models, including that of the Band function fits.}%\begin{scriptsize}
% \vspace{0.3in}
\begin{tabular}{cccccccccc}

\hline
Bin \# & Interval (s) & Model & $kT_{\rm h}$\,(keV) &  $N_{\rm h}$\,($10^{-1}$) & $kT_{\rm l}$ or $kT$\,(keV) & $N_{\rm l}$\,($10^{-1}$) & $\Gamma$ & $N_{\Gamma}$\,($10^{-1}$) & $\chi^2$\,(dof) \\
\hline
\hline
$15$ &$88.2-94.2$     & PL    &                             &                             &                             &                          & $  1.39_{ -0.05}^{+  0.05}$ & $  5.4_{ -0.4}^{+  0.5}$ & $ 113.4\,(75)$ \\
     &                & BBPL  &                             &                             & $  12_{ -2}^{+  2}$         & $  2.1_{ -0.3}^{+  0.3}$ & $  1.97_{ -0.12}^{+  0.13}$ & $  8.4_{ -0.8}^{+  0.8}$ & $  86.9\,(73)$ \\
     &                & 2BBPL & $  38_{ -13}^{+  52}$ & $  3.8_{ -1.4}^{+ 13}$            & $  6_{ -1}^{+  1}$          & $  1.4_{ -0.3}^{+  0.3}$ & $  2.21_{ -0.16}^{+  0.17}$ & $  8.9_{ -0.8}^{+  0.8}$ & $  70.8\,(71)$ \\
&&Band&&&&&&&103.1 (73)\\
     \hline
$16$ &$94.2-104.2$    & PL    &                             &                             &                             &                          & $  1.40_{ -0.04}^{+  0.04}$ & $  4.6_{ -0.3}^{+  0.3}$ & $  88.9\,(75)$ \\
     &                & BBPL  &                             &                             & $  11_{ -2}^{+  2}$         & $  1.1_{ -0.3}^{+  0.3}$ & $  1.64_{ -0.10}^{+  0.11}$ & $  5.4_{ -0.5}^{+  0.5}$ & $  77.8\,(73)$ \\
     &                & 2BBPL & $  17_{  -3}^{+   4}$ & $  1.6_{ -0.3}^{+  0.3}$          & $  4_{ -1}^{+  1}$          & $  0.8_{ -0.2}^{+  0.2}$ & $  2.14_{ -0.22}^{+  0.23}$ & $  6.2_{ -0.5}^{+  0.5}$ & $  70.4\,(71)$ \\
&&Band&&&&&&&82.8 (73)\\
\hline
$17$ &$104.2-137.2$   & PL    &                             &                             &                             &                          & $  1.43_{ -0.03}^{+  0.04}$ & $  3.0_{ -0.1}^{+  0.1}$ & $  89.5\,(75)$ \\
     &                & BBPL  &                             &                             & $  14_{ -5}^{+  8}$         & $  0.4_{ -0.2}^{+  0.2}$ & $  1.55_{ -0.07}^{+  0.07}$ & $  3.2_{ -0.2}^{+  0.2}$ & $  84.9\,(73)$ \\
     &                & 2BBPL & $  25_{  -5}^{+  10}$ & $  1.1_{ -0.3}^{+  0.5}$          & $  4.3_{ -0.4}^{+  0.6}$    & $  0.5_{ -0.1}^{+  0.1}$ & $  1.93_{ -0.14}^{+  0.15}$ & $  3.6_{ -0.2}^{+  0.2}$ & $  72.8\,(71)$ \\
&&Band&&&&&&&77.8 (73)\\
\hline
$18$ &$137.2-253.2$   & PL    &                             &                             &                             &                          & $  1.38_{ -0.04}^{+  0.04}$ & $  0.92_{ -0.04}^{+  0.04}$ & $  96.8\,(75)$ \\
     &                & BBPL  &                             &                             & $  7_{ -1}^{+  1}$          & $  0.3_{ -0.1}^{+  0.1}$ & $  1.59_{ -0.08}^{+  0.09}$ & $  1.00_{ -0.06}^{+  0.06}$ & $  75.8\,(73)$ \\
     &                & 2BBPL & $   7_{  -1}^{+   1}$ & $  0.2_{ -0.1}^{+  0.1}$          & $  0.07_{ -0.01}^{+  0.02}$ & $  0.2_{ -0.1}^{+  0.3}$ & $  1.51_{ -0.07}^{+  0.08}$ & $  0.93_{ -0.06}^{+  0.06}$ & $  65.8\,(71)$ \\
&&Band&&&&&&& $96.2 (73)$\\
\hline
$19$ &$253.2-573.2$   & PL    &                             &                             &                             &                             & $  1.50_{ -0.06}^{+  0.06}$ & $  0.03_{ -0.02}^{+  0.02}$ & $  14.3\,(17)$ \\
\hline

\end{tabular}

\label{t_late_2bbpl}
% \end{scriptsize}
\end{table*}

The BAT and XRT spectrum at late times are extracted 
when the S/N of the \fe~ does not allow any further time division. The XRT data in WT mode
and the BAT data are available until 570\,s and $\sim250$\,s, respectively. Note that the 
XRT light curve has two breaks one at 91\,s and another at 134\,s, both counted with respect 
to the BAT trigger time. Hence, in choosing the time intervals, we respect these break times 
and choose roughly logarithmic bins i.e., the interval length increases roughly as a
geometric progression. We obtain 5 bins: 85--91\,s, 91--101\,s, 101--134\,s, 134--250\,s
and 250--570\,s (from BAT trigger time). The simultaneous BAT data is extracted in the first 
four intervals.

The time-resolved data is analysed with a power law (PL), BB + PL, 2BB + PL and Band function. As 
before, two absorption terms --- the Galactic $n_{\rm H}$ is set to 
$7.74\times10^{20}$\,atoms cm$^{-2}$ and the source $n_{\rm H}$ to $5.1\times10^{21}$\,atoms cm$^{-2}$, are used.
The result of the fits are shown in Table~\ref{t_late_2bbpl}. We find that the PL 
does not give a good fit (see Table~\ref{t_late_2bbpl}) and shows large residual. The spectra are much better fitted with 
the BB+ PL model and even better with the 2BB + PL model. This shows that there is
a curvature in the spectrum. Hence, this phase most probably corresponds to the prompt 
emission phase. We found that the Band function shows an equally bad fit as the PL model 
and the $E_{\rm p}$ could not be constrainred. The 2BB + PL model seems to be the best fit at the 
late time. For the last bin, as the BAT data is no more available we could not constrain the 
parameters of more complicated models other than the PL in the limited bandwidth of the XRT.
Hence, only the results of PL fit is reported for this bin. 

In Fig.~\ref{2bbpl_late_evolution}, we show the evolution of the parameters of the 
2BB + PL model taking all the observations together. The panels are same as in Fig.~\ref{2bbpl_parm}
that refers to the 2BB + PL model fitted at early times. In Fig.~\ref{babb_late_evolution}, 
similar plot is shown for the Band + BB model. In both cases, the evolution at the late phase 
is consistent with the previous evolution. The late time spectral evolution together with a 
significantly better reduced $\chi^2$ in the high resolution data of the XRT 
suggests that the 2BB + PL model probably most consistently captures the 
overall spectral evolution of the burst.

% In Panel 3, the flux evolution clearly shows that 
% at the late phase the thermal flux rapidly decreases while that of the non-thermal component
% lingers at the late phase. In \cite{Basak_Rao_2013_linger}, it was shown that for the GRBs with 
% high GeV emission the non-thermal component shows this behaviour at late phase. The current observation
% is consistent with that finding.

\subsection{Evidence for a second hard pulse}\label{second_pulse}

\begin{figure}\centering
{
\includegraphics[width=\columnwidth]{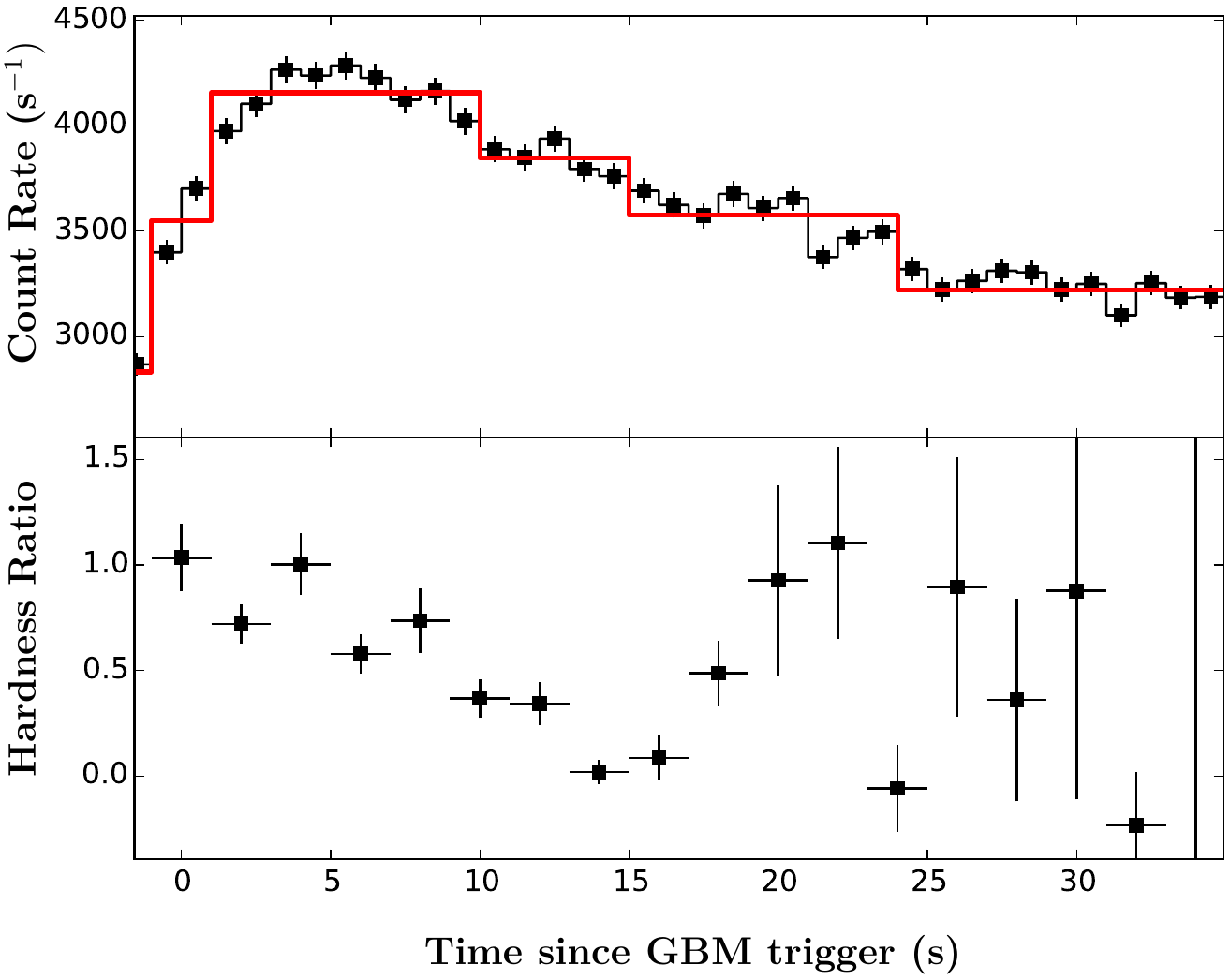} 
}
\caption{\emph{Upper panel:} Bayesian blocks shown on the light curve of combined NaI 0, 1 and 3 in 8--900\,keV.
The blocks are -1--1\,s, 1--10\,s, 10--15\,s and 15--24\,s. clearly, a new block starts near the time of first 
detected LAT photon at $\sim 18$\,s. \emph{Lower panel:} Hardness ratio defined as the ratio of count between 
500\,keV--5\,MeV and 200\,keV--400\,keV band of BGO 0. A sudden increment of hardness is apparent at $\sim 18$\,s.
}
\label{lc_hr}
\end{figure}

In case of GRBs with multiple but separable pulses, it is frequently observed that the first pulse shows 
a HTS evolution, followed by a jump in the peak energy during the onset of a second pulse,
and then again showing a HTS evolution in the falling part of that pulse (see e.g., 
\citealt{Ghirlandaetal_2010, Luetal_2012}). 
A very similar behaviour is seen for GRB 151006A, though the presence of a second pulse is not 
seen in the lightcurve. The late time spectral evolution being consistently HTS since the time of sudden jump
($\gtrsim 18$\,s, see Section~\ref{late_phase}) indicates that the second phase is probably a second pulse 
of the prompt emission hidden in the data. This is not readily evident probably because it is expected to be a hard pulse, 
where the signal is weak. In order to have a better look at the phenomenon we perform the following analysis. 

We first construct the Bayesian blocks for the combined count rate data of NaI 0, 1 and 3 in 8--900\,keV
(see Fig.~\ref{lc_hr}, upper panel). The Bayesian block approach finds the best possible way to represent a 
time-series data as a series of blocks or segments such that the signal underlying each block is 
constant within the observational error (\citealt{scargel2013}). We use the dynamical programing 
algorithm of \cite{scargel2013} to construct these blocks which are then over-plotted on the 
count rate lightcurve in Fig.~\ref{lc_hr}, upper panel. During the interval shown in the Figure, 
we find the Bayesian blocks as -1--1\,s, 1--10\,s, 10--15\,s and 15--24\,s. With only the NaI 3 
detector these blocks are -1--1\,s, 1--14\,s and 14-24\,s. We note that a new block starts 
at $\sim14-15$\,s which shows that the count rate flux of this bin is statistically different from the 
Bayesian blocks on either side on the time axis. But, as the flux level of the consecutive blocks 
around this block decreases monotonically, there is no evidence for a new pulse in the data. 
However, we note from Table~\ref{t_bandbb} that the peak energy at the transition time reaches a very 
high value $\sim 5$\,MeV, and hence, it is unlikely that such a change will give rise to any
significant pulse profile in the NaI detectors. On the other hand, the flux level of BGO 
detector is quite poor to carry out such analysis. 

Given the high value of the peak energy, and no pulse profile with Bayesian blocks in the NaI detector,
we are left with only one possibility that the pulse is hidden in the BGO data and we investigate 
that possibility. Before proceeding, we note that there are two competing factors here:
a lightcurve constructed in a wide energy 
band will smear out the small variations that we expect at the high energies. On the other hand, a 
lightcurve in a limited bandwidth suffers from the poor statistics of photon count, more so 
for the BGO detector and that too at a time when the photon flux is already low. This is why we used 
the full band of the NaI detector for the Bayesian block analysis in the first place.
A more robust way to investigate it further is to study the evolution of hardness ratio (HR)
rather than the count rate lightcurve in a limited band. As the HR is a ratio of count between 
two energy bands, the small changes in the photon count subjected to the change in spectral 
peak will be amplified. More importantly, if there is a smooth pulse presumably a hard one, which is 
not seen in the lightcurve otherwise, the HR should track that pulse profile. We use the BGO 
detector (the BGO 0) rather than the NaI since the peak energy reaches high values covered by the BGO energy band.
On the other hand, we cannot use the LAT LLE data as it does not provide the low energy band required for this analysis.
We choose the hard band as $H=\rm 500\,keV-5\,MeV$ and the soft band as $S=\rm 200\,keV-400\,keV$,
and then the Hardness Ratio $=H/S$. 
% The definition is chosen such that 
% the ratio becomes close to 1, and hence, does not suffer due to the uncertainty of 
% very low count in one band. 
We then stick to the same definition throughout.
The evolution is shown in Fig.~\ref{lc_hr}, lower panel.
The first phase shows a HTS evolution starting with a hardness ratio of $\sim1$, 
reaching down to a value close to 0 until $\sim15$\,s and then shows a sudden jump again 
reaching a hardness ratio of $\sim1$, followed by another HTS evolution. The errors 
in the hardness ratio are larger at the later phase due to the lower flux level, but the evidence 
of the jump in hardness lightcurve, and a smooth pulse profile afterwards is very clear. 
This changeover time remarkably coincides with the onset of the unusual spectral evolution.
% the new Bayesian block at this time. 
This is exactly when the second phase of emission begins 
and the first LAT photon ($>100$\,MeV) is detected. This analysis, combined with the fact that the 
second phase shows a very similar HTS evolution as seen in other GRBs, points towards 
the fact that this new energy injection is due to a second hard pulse of the prompt emission 
itself.

\begin{figure}\centering
{
\includegraphics[width=\columnwidth]{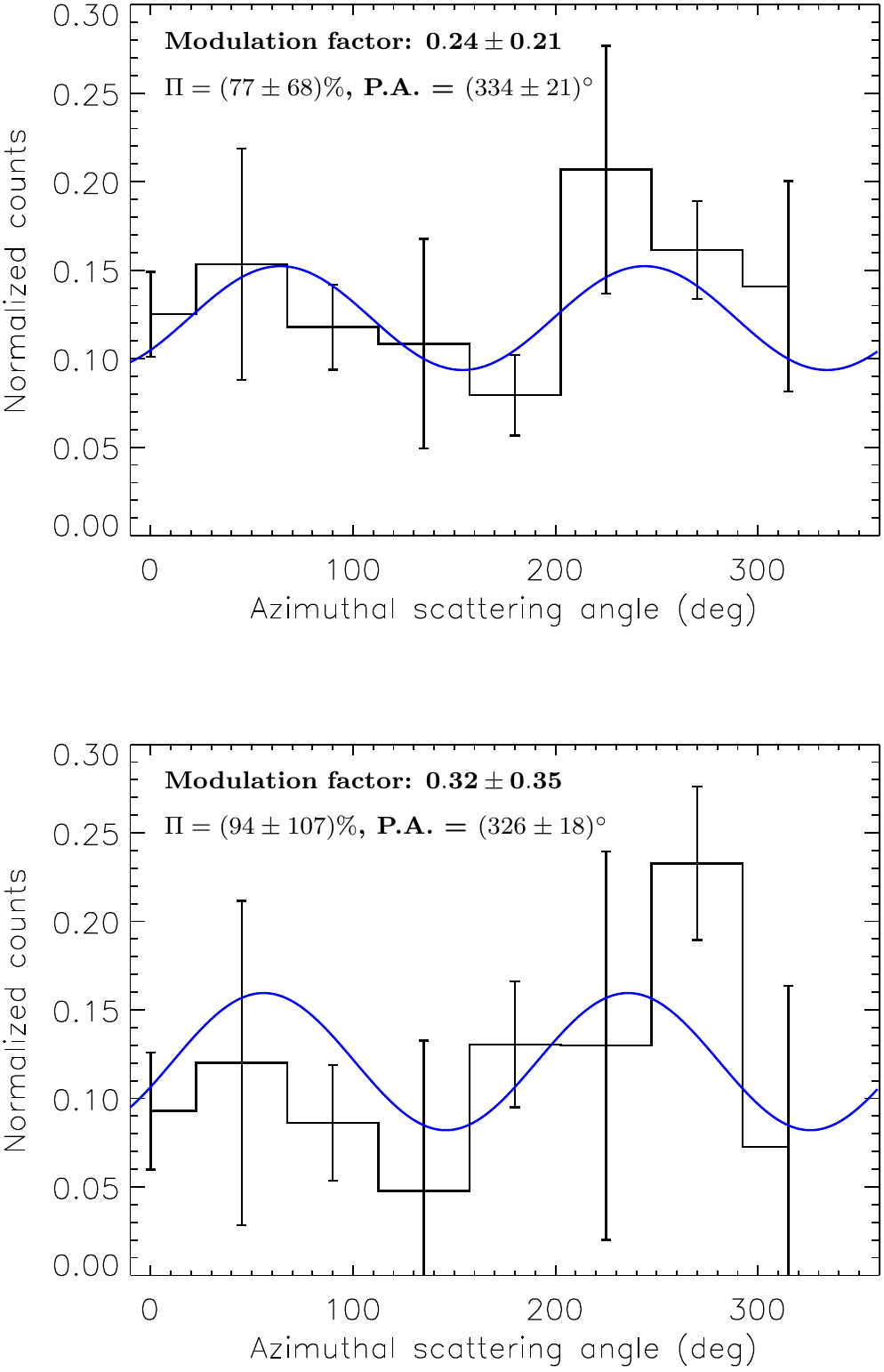} 
}
\caption{Modulation curves for GRB 151006A in two phases: 0--16\,s (upper panel) and 
16--33\,s (lower panel). The modulation curve is obtained
after geometrically correcting the raw azimuthal angle histogram by 
normalizing with respect to a simulated unpolarized distribution for a 
similar spectra and off-axis detection angle (\citealt{Raoetal_2016}). The blue solid line
is the $\cos{\phi}^2$ fit to the modulation curve. The fitted modulation
factor, polarization fraction ($\Pi$) and polarization angle (P. A.) are 
shown in the inset of the respective panels, with errors 
estimated at 68\% confidence level (see text for details).
}
\label{pol}
\end{figure}

\subsection{Polarization measurement -- further evidence for the second pulse of prompt emission}

Another piece of information about these two emission phases can be obtained from 
polarization measurement. If the second emission phase is a hard pulse, and we attribute 
polarization to the non thermal processes that produce harder spectrum, then we expect to have 
an enhancement of the degree of polarization in the second phase.
CZTI being a pixelated detector and having a significant Compton scattering 
probability at energies beyond 100 keV, essentially works as a 
Compton polarimeter at 
these energies. The double pixel events within the photon tagging time
window of 40 $\mu$s which satisfy the Compton kinematics are identified
as valid Compton events. The validity of the Compton event selection 
and the polarization measurement capability of CZTI have been established
by detailed simulation and experimental studies during ground calibration
of CZTI (\citealt{chattopadhyay14,Vadawale_etal_2015}). The first onboard validation 
of CZTI polarimetry was obtained with the detection of polarization of
GRB 160131A (\citealt{vadawale16}) and GRB 151006A (\citealt{Raoetal_2016}). GRB151006A, though
is moderately bright for X-ray polarization measurement, a hint of 
polarization is seen with an estimated polarization degree $> 90\%$ with
a detection significance of 1.5$\sigma$ (68$\%$ confidence level with 
1 parameter of interest). Since our spectroscopic analysis shows that there
could be two distinct phases of emission, we tried to explore polarization
measurement in these two sectors using the CZTI data. Fig.~\ref{pol} shows the
modulation curves for these two time bins in 100 $-$ 350 keV, where the blue
solid line is the sinusoidal fit to the observed distribution of 
azimuthal angle of scattering.

In order to estimate the polarization fractions, we did detailed 
Geant 4 simulations (\citealt{agostinelli03}) using \emph{Astrosat} mass model 
which includes all the instruments of \emph{Astrosat} along with the complete satellite structure. We employ
the fitted spectroscopic models to simulate the energy distribution of the
incident photons in Geant 4 to obtain the modulation factors for 
100$\%$ polarized beam ($\mu_{100}$) which are then used to estimate the polarization fractions
($P = \mu/\mu_{100}$) in two phases, 0--16\,s and 16--32\,s. Estimated polarization
fractions are 77$\%$ and 94$\%$ with $1\sigma$ detection significance
at polarization angles 334$^\circ$ and 325$^\circ$ respectively in these 
two phases. The detection significance of polarization
has reduced significantly compared to the initial report of
polarization for GRB 151006A (\citealt{Raoetal_2016}) which is
expected due to reduced number of events in the individual phases. Though
fraction and angle of polarization are poorly constrained in both the time 
sectors, the measured values do not show any decrement of polarization.
In the multi-component models with thermal and non thermal parts, it is 
suggested that the thermal emission dominates in the first part while the 
non thermal processes become important at the later stage (e.g., \citealt{Gonzalezetal_2003}).
Attributing the polarization to the non thermal processes,
the above evolution is consistent with the expected behaviour for such models. 
Hence, there is a hint of multi-component spectra in the data.

It is important to note that the polarization degree in the afterglow phase found 
by optical measurements so far show a pretty low value ($\lesssim10\%$) and it is 
expected to reduce further as the afterglow proceeds,
see e.g., \cite{greiner2003, mundell2007}. This appears to be due to the fact that 
the magnetic field during the afterglow phase is mostly generated by turbulence 
and therefore has a random orientation, having only a small coherence length. This is in contrast 
with the prompt emission phase, where measurements show a high degree of polarization ($\sim40-80\%$) which 
can be achieved by an ordered magnetic field, see \cite{Waxman2003}. Though a different 
mechanism is possible to achieve a high degree of polarization of the prompt emission,
the important point is that such high values are not seen in the afterglow phase.
Our measurement is consistent with a high value, and though the statistics is 
poor the measured values in the two phase do not contradict the interpretation
that we are most probably observing the prompt emission extending out to the second phase. Hence, the 
data is indicative of a second hard pulse rather than the onset of an afterglow phase.

\subsection{Comparison between the models}

\begin{figure}\centering
{\vspace{-0.1in}
\includegraphics[width=\columnwidth]{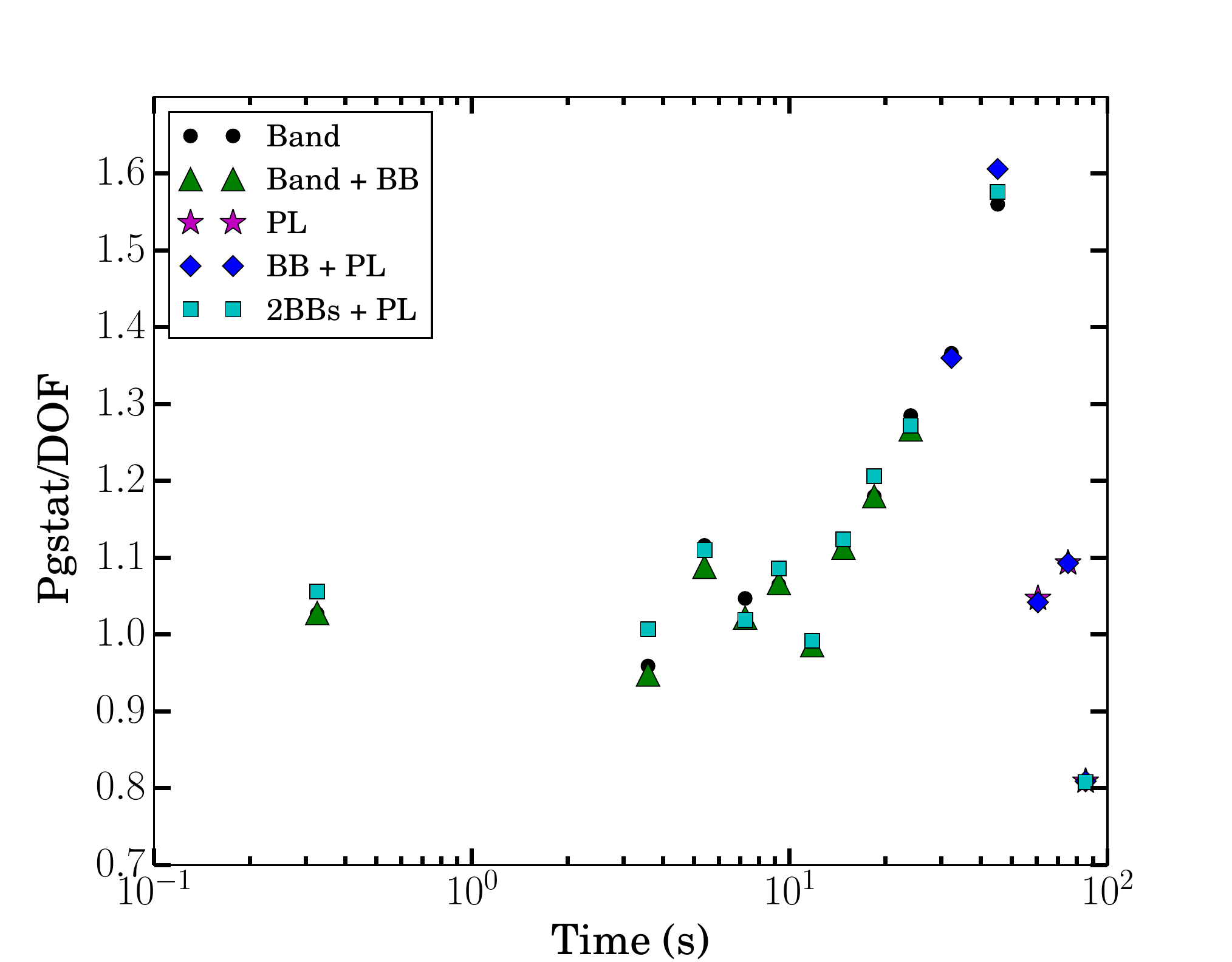} 
}
\caption{The PGStat/dof of all the models are shown.
}
\label{pgstat}
\end{figure}

\begin{figure}\centering
{
\includegraphics[width=\columnwidth]{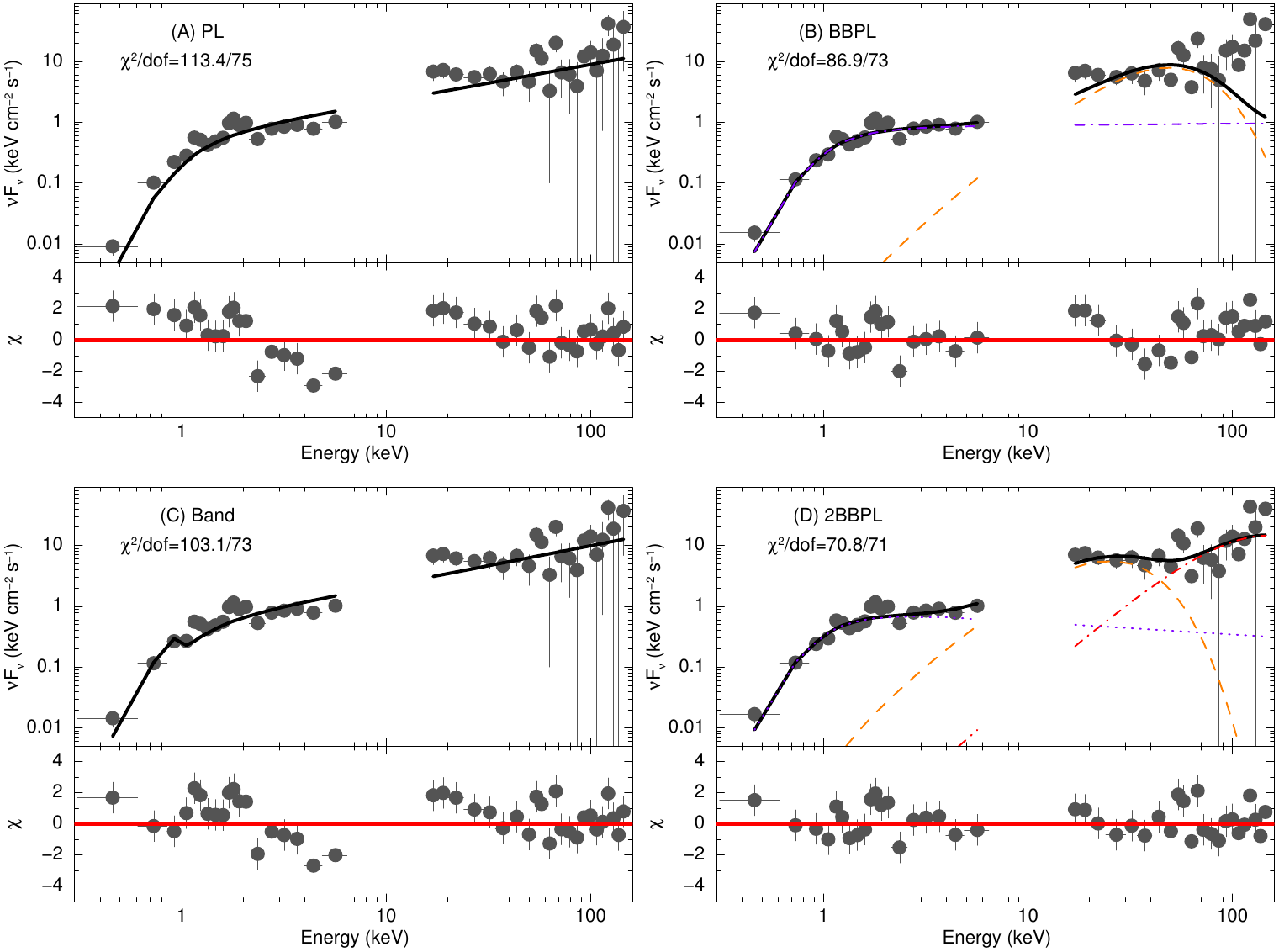} 
}
\caption{Spectral fitting of the joint BAT and XRT data with PL, BB + PL, Band and 2BB + PL models 
for the late time bin 15. 
The data and residual are shown by grey filled circles. The high energy blackbody, 
low energy blackbody (or the blackbody for the BB + PL model), the power law and
the total model, are shown by red dot-dashed line, orange dashed line, violet dotted
line and thick black line, respectively.
}
\label{late_spectrum}
\end{figure}

We then compare the different models based on their goodness of fit.
The PG-Stat/dof of all the models in different time bins are shown in Fig.~\ref{pgstat}.
We note that all the models have very similar PG-Stat. Thus, making the judgment of the 
best model is quite indecisive. We then use Bayesian inference criteria (BIC) to see 
whether the more complicated models are indeed required by statistics 
(e.g., \citealt{wang2017}). The BIC 
is defined as $-2\ln \cal{L}$$+ k\ln(\nu+k)$, where $-2\ln \cal{L}$ is log likelihood,
$k$ is the number of free parameters and $\nu$ is the degrees of freedom. In Table~\ref{bic}, 
we show the BIC values and the preferred models.

\begin{table}\centering
\caption{BIC values and preferred model}%\begin{scriptsize}
% \vspace{0.3in}
\begin{tabular}{ccccc}

\hline
\#  &  \multicolumn{4}{c}{$-2\ln \cal{L}$ (BIC)}\\
        & Band & Band+BB & 2BB+PL & Preferred \\
\hline
\hline
1  & 489.8 (514.5)   & 488.1 (525.1)   & 501.7 (538.7)   & Band \\
2  & 457.2 (481.9)   & 450.4 (525.1)   & 478.3 (515.3)   & Band \\
3  & 532.3 (557.0)   & 517.0 (554.0)   & 527.4 (564.4)   & Band+BB \\
4  & 499.2 (523.9)   & 485.5 (522.5)   & 483.8 (520.8)   & 2BB+PL \\
5  & 507.8 (532.5)   & 506.7 (543.7)   & 515.8 (552.8)   & Band \\
6  & 471.2 (495.9)   & 468.3 (505.3)   & 471.1 (508.1)   & Band \\
7  & 529.5 (554.2)   & 528.6 (565.6)   & 534.1 (571.1)   & Band \\
8  & 574.9 (599.7)   & 572.5 (609.7)   & 578.1 (615.3)   & Band \\
9  & 612.8 (637.5)   & 601.9 (638.9)   & 604.2 (641.2)   & Band \\
10 & 640.5 (665.1)   &                 &                 & Band \\
11 & 747.4 (772.1)   &                 & 751.1 (788.2)   & Band \\
\hline
   & BBPL            & BBCPL           & PL \\
   \hline
12 & 1497 (1526) & 1498 (1534) & 1505 (1520) & PL \\
13 & 1571 (1600) & 1566 (1602) & 1573 (1587) & PL \\
14 & 1162 (1191) & 1162 (1199) & 1165 (1179) & PL \\
\hline
   & BBPL            & 2BB+PL          & PL \\
   \hline
15 & 86.9 (104.3)    & 70.8 (96.9)     & 113.4 (122.1)   & 2BB+PL \\
16 & 77.8 (95.2)     & 70.4 (96.5)     & 88.9 (97.6)     & BBPL/2BBPL \\
17 & 84.9 (102.3)    & 72.8 (98.9)     & 89.5 (98.2)     & PL/2BBPL \\
18 & 75.8 (93.2)     & 65.8 (91.9)     & 96.8 (105.5)    & 2BB+PL \\ 
\hline
\end{tabular}
\label{bic}
% \end{scriptsize}
\end{table}

We note that in the initial bins where we use the \fe~ data, the Band function is the
preferred model. This signifies that the data is consistent with the Band function 
and a more complicated model is not statistically required. There are only two exceptions,
Bin 3 and 4. In the former Band + BB is the preferred model, while in the later the 
2BB + PL is the preferred model. At the later phase, we see even the PL is the 
preferred model and the other complicated models are not statistically required. 
Our result clearly shows the limitations of a background dominated low resolution 
detector. 

However, in the late phase, though we have lower count rate, the signal to noise of
the data is improved due to both BAT and XRT and we have good resolution of the XRT. 
We see a preference for the 2BBPL model in the data at late phase.
From Table~\ref{t_late_2bbpl}, we also see that the Band function is not 
preferred. In fact, we could not constrain the peak energy, and it signifies that the 
spectral shape is not a Band function.

In Fig.~\ref{late_spectrum}, we show the spectral fit and residuals of the PL, BB + PL, Band and 2BB + PL model fit
for bin 15. Note that for the PL fit the slope at lower and higher energies do not match which is 
clearly indicated by the deviation of the residual in the opposite direction of the zero line.
For the BBPL model, this is improved, but then we find large residual on the both sides of the 
blackbody peak. The Band function fares no better than the PL model and has a similar residual.
All the curvatures in the spectrum are consistently taken care by the 2BB + PL model.

\section{Conclusions and Discussion}
\label{disc}

% \subsection{Two phases of emission}

Single pulse bursts have been carefully studied using time resolved spectroscopy, 
as it is generally assumed to alleviate the issue of pulse overlapping which hinders 
our understanding of spectral evolution of GRB with time. GRB 151006A  is a single
pulse burst in different energy bands and thereby is an ideal 
one for time resolved spectroscopy. In general, such bursts exhibit a HTS
spectral peak evolution. However, the current analysis of GRB 151006A 
shows an unexpected behaviour of change in trend from HTS
evolution to increasing spectral peak with time, after a few seconds from 
the burst trigger. Coincidently, this rise is observed after the arrival 
of LAT  photons ($> 100 \, \rm MeV$). Such a dramatic change in the spectral
behaviour for single pulse burst is observed for the first time. 
This new injection of energy does not show any additional pulse
in the counts light curve. Thus, this cautions us that a single 
pulse need not always suggest a HTS spectral peak evolution and
that there may be other pulses hidden in the light curve profile.

It is important to notice that high energy emission is often delayed 
and generaly modelled with powerlaw appearing in the spectrum at late times
(e.g., \citealt{Gonzalezetal_2003, Ackermannetal_2013_LAT}). This can potentially 
change the behaviour of the spectral evolution we report here. However, we have 
tested that a powerlaw with Band function is not statistically required
for Bin 8 and 11. This is apparent from the spectral shape in Fig.~\ref{wide_spectrum}.
GRB 151006A is not among the high-fluence LAT class  
(cf. \citealt{Ackermannetal_2013_LAT}), and for several such cases the spectrum 
is found to be well fitted with Band function only and no powerlaw is required.

In order to show that the $E_{\rm p}$ variation is physical and not due 
to single peak Band function, we also tried models with two peaks i.e., Band + BB 
and 2BB + PL, and re-confirm the spectral variation.
The analysis also presents one of the key issues in current GRB spectral 
analysis i.e., a preferred model cannot be decided based only on statistics.
Though in the later part the 2BB + PL model seems to be a better fit.
In \cite{Basak_Rao_2015_090618} the spectral components of this model is proposed to
originate in a spine-sheath jet (\citealt{Ramirez-Ruizetal_2002, Zhangetal_2003, Zhangetal_2004},
see \citealt{Iyyanietal_2015} for an alternative explanation).
The blackbody radiations are produced at the two photospheres.
In addition, photons crossing the boundary layer of the spine-sheath structure are 
on an average Compton up-scattered and hence form a high energy power-law spectrum. 
Further non thermal processes can occur due to 
synchrotron emission at a higher radial distance. These photons are naturally delayed
with respect to the photons produced by thermal and non thermal processes stated above. 
Hence, the second phase in this scenario is probably an onset of the delayed emission 
phase. We note here that though the second phase starts after $\sim18$\,s, the pulse 
can have a much lower start time, which is hidden in the falling part of the first pulse.
Hence, an actual delay of this phase can be much smaller, though remains undetermined.

Recently, \cite{Moretti&Axelsson2016} reported a rise in the high energy spectral peak with time, 
for the first LAT detected GRB 080825C, a multi-pulse burst (\citealt{Abdo_etal_2009_080825C}), 
when it was re-analysed including the LLE data. 
% Even though they find a different combination of
% functions to model the spectrum, it was for the first time such an increasing trend of spectral
% peak energy had been reported. 
This points out the significance of LLE emission 
detections which can be crucial in constraining the high energy 
spectral behaviour of GRB spectra. The interesting part of our 
observation is that the increasing beahaviour is seen in an otherwise 
single pulse. The new injection of energy
may be associated with a second pulse of emission in the prompt
phase which brings about a dynamical change in the behaviour of
the radiation process, or it may be an overlap of the onset of 
afterglow (emission from a different region) with the prompt 
emission, which then significantly dominates the high energy 
and thereby making the prompt emission indiscernible in the spectrum. 
We note that the $\alpha$ values of Band function to be consistent 
with slow cooling synchrotron radiation in the beginning of the burst ($< -0.67$) 
and gets softer with time approaching the fast cooling index 
($\alpha = -1.5$). If the new energy injection is due to a second pulse 
in the prompt emission, we may be observing a transition in the microphysical
parameters related to the synchrotron radiation, such as strength or structure 
of magnetic fields, fraction of energy injected into electrons, radiation
efficiency etc (\cite{Daigne&Mochkovitch1998}), causing the transformation 
from slow to fast cooling of electrons, see \cite{Iyyani_etal_2016}. 
Since this transformation is observed to be sudden, it perhaps indicate the
onset of a second pulse with a different source parameter.
For the afterglow scenario,
such emission has been mostly modelled using a power law
(\citealt{Ackermannetal_2013_LAT, Kumar2009, Ghisellini_etal_2010}).
However, we note a definite spectral curvature at late times and 
the detection of LAT emission 
is less significant and not many photons are detected.
Our analysis with Bayesian block and evolution study of the hardness 
ratio also points towards the former scenario i.e., a new hard pulse hidden in the data.

% \subsection{Physical Interpretation of the spectral models}

% \subsection{Polarization: present data and future perspective}
Finally, the inability of determining the best model based on statistics 
also summons for constraining observations like polarization. This can potentially
characterize various radiation processes leading to the observed emission
as well as in revealing the structure and strength of magnetic fields of emitting region.
This can in turn be an invaluable input in enhancing our understanding of 
shock physics as well as the content of GRB jets.  
A significant polarization detection requires high photon
statistics and lack of this has actually prevented the CZTI instrument onboard 
\emph{Astrosat} in constraining the polarization measurement for GRB 151006A. 
The current measurement hints a moderately high polarization, however, with low 
significance, and thereby is consistent with nearly all model predictions.
However, the capability demonstrated by CZTI, offers a promising era of 
polarization detections above $100 \, \rm keV$ and also its time dependent study, in case of brighter GRBs.

\section*{Acknowledgments} 
We gratefully acknowledge the referee for comments on reshaping 
the paper. This publication uses data from the \emph{Astrosat} mission of the
Indian Space Research Organisation (ISRO), archived at the
Indian Space Science Data Centre (ISSDC). CZT-Imager is built
by a consortium of Institutes across India including Tata Institute
of Fundamental Research, Mumbai, Vikram Sarabhai Space
Centre, Thiruvananthapuram, ISRO Satellite Centre, Bengaluru,
Inter University Centre for Astronomy and Astrophysics, Pune,
Physical Research Laboratory, Ahmedabad, Space Application
Centre, Ahmedabad: contributions from the vast technical team
from all these institutes are gratefully acknowledged.
This research has made use of data obtained through the
HEASARC Online Service, provided by the NASA/GSFC, in support of NASA High Energy
Astrophysics Programs. RB is a stipendiary of START program of the Polish 
Science Foundation (2016) and supported by 
Polish National Science Centre grants 
2013/08/A/ST9/00795,
2013/10/M/ST9/00729 and
2015/18/A/ST9/00746.

\appendix

\section{Table of spectral analysis}
\label{table_spectral}

Here we have reported the parameters obtained from time resolved spectral analysis using the Band + BB (Table~\ref{t_bandbb}),
2BB+PL (Table~\ref{t_2bbpl}) and BB+PL (Table~\ref{t_bbpl}). 
In Table \ref{t_bbpl}, we also show the Pg-stat/dof of two more models i.e power law (PL) and
blackbody + cutoff power law (BB + CPL), for comparison.

%These are powerlaw (PL) and BB+Cutoff powerlaw (BBCPL). 
% The former is a simpler model than BB + PL while the 
% latter adds a curvature at the higher energy. We note that compared to these models, the BB + PL is 
% fairly acceptable. 

\begin{table*}\centering
\caption{Parameters of time resolved spectral fitting with Band + BB model in bin 1--9 }%\begin{scriptsize}
% \vspace{0.3in}
\begin{tabular}{ccccccccc}

\hline
Bin \# & Time interval (s) & $kT$\,(keV) &  $N_{\rm BB}$ & $\alpha$\,(keV) & $\beta$ & $E_{peak}$\,(keV) & $N_{\rm Band}\,(10^{-3})$ & PG-Stat (dof)\\
\hline
\hline
$1$       &$-2.0$ -- $ 2.7$ & $  22_{ - 8}^{+ 22}$  & $   0.12_{  -0.08}^{+   0.08}$  & $   -0.85_{ - 0.05}^{+   0.06}$  & $   -2.8_{  -0.2}^{+    0.2}$  & $    3220_{  -2294}^{+  770 }$  & $  3.1_{  -0.2}^{+   0.2}$  &  $  488.1\,(475)$\\ 
$2$       &$ 2.7$ -- $ 4.5$ & $  14_{ - 2}^{+2}$  & $  0.61_{  -0.11}^{+   0.11}$  & $  -1.05_{  -0.04}^{+   0.06}$  & $   -2.6_{  -0.2}^{+   0.2}$  & $   2892_{  -1919}^{+ 7099}$  & $   8.9_{  -0.4}^{+  0.4}$  &  $  450.4\,(475)$\\
$3$       &$ 4.5$ -- $ 6.3$ & $  23_{ -2}^{+  2}$  & $  1.22_{  -0.16}^{+   0.15}$  & $   -1.12_{  - 0.06}^{+  0.05}$  & $   -2.9_{  -0.4}^{+  0.2}$  & $  2281_{  - 699}^{+  7709}$  & $  8.3_{  -0.4}^{+  0.4}$  &  $ 517.0\,(475)$\\
$4$       &$ 6.3$ -- $ 8.2$ & $   28_{-2.}^{+  3}$  & $  1.42 _{  -0.16}^{+   0.20}$  & $  -1.34_{  - 0.03}^{+    0.06 }$  & $   -2.8_{  - 0.6}^{+  0.3}$  & $   4893_{  -2717}^{+  2193}$  & $  7.8_{  -0.3}^{+  0.5}$  &  $  485.5\,(475)$\\
$5$       &$ 8.2$ -- $10.3$ & $ 12_{ - 4}^{+ 6}$  & $    0.25_{  -0.16}^{+  0.23}$  & $   -1.16_{  -0.04}^{+    0.14}$  & $  -2.4_{  - 0.1}^{+    0.1}$  & $   986_{  -366}^{+    1135 }$  & $   10.3_{  - 0.5}^{+  1.1}$  &  $  506.7\,(475)$\\
$6$       &$10.3$ -- $13.2$ & $   12_{ -2}^{+  4}$  & $    0.34_{  -0.20}^{+   0.21}$  & $  -1.10_{  -0.13}^{+   0.17}$  & $   -2.4 _{  - 0.2}^{+    0.1}$  & $   610_{  -221}^{+  541}$  & $  8.8_{  -0.1}^{+   0.1}$  &  $ 468.3\,(475)$\\
$7$       &$13.2$ -- $16.3$ & $  10_{ -\infty}^{+  20}$  & $   0.19_{  -\infty}^{+   0.20}$  & $  -1.09_{  -0.18}^{+   0.22}$  & $ -2.3_{  -0.2}^{+  0.1}$  & $  412_{  -158}^{+  374 }$  & $  9.6_{  -0.2}^{+  0.2}$  &  $ 528.6\,(475)$\\
$8^{(a)}$ &$16.3$ -- $20.6$ & $ 33_{ -4}^{+  4}$  & $    0.74_{  -0.10}^{+   0.12}$  & $    -1.50_{  - 0.03}^{+    0.05}$  & $  -3.0_{  -0.9}^{+    0.2}$  & $  5274_{  -4511}^{+  3102 }$  & $  4.2_{  - 0.1}^{+    0.3}$  &  $ 572.5\,(485)$\\
$9$       &$20.6$ -- $27.5$ & $  19_{-3}^{+ 3}$  & $   0.39_{  -0.06}^{+   0.06}$  & $ -1.42_{  -0.03}^{+   0.03}$  & $   -3.5_{  -\infty}^{+   0.7}$  & $   5078_{  -2228}^{+  2370}$  & $  3.1_{  -0.2}^{+   0.2}$  &  $ 601.9\,(475)$\\

\hline
\end{tabular}

Note: (a) Including LAT data
\label{t_bandbb}
% \end{scriptsize}
\end{table*}

% \clearpage

\begin{table*}\centering
\caption{Parameters of time resolved spectral fitting with 2BB + PL model}%\begin{scriptsize}
% \vspace{0.3in}
\begin{tabular}{ccccccccc}

\hline
Bin \# & Time interval (s) & $kT_{\rm h}$\,(keV) &  $N_{\rm h}$ & $kT_{\rm l}$\,(keV) & $N_{\rm l}$ & $\Gamma$ & $N_{\Gamma}$ & PG-Stat (dof)\\
\hline
\hline
$1$        &$-2.0$ -- $ 2.7$ & $ 730_{ -91}^{+ 103}$  & $  23_{  -4}^{+   4}$  & $  89_{ -16}^{+  18}$  & $   2.6_{  -0.6}^{+   0.8}$  & $  1.62_{ -0.03}^{+  0.03}$  & $   2_{  -1}^{+   1}$  &  $ 501.7\,(475)$\\
$2$        &$ 2.7$ -- $ 4.5$ & $ 302_{ -59}^{+  70}$  & $  19_{  -4}^{+   4}$  & $  36_{  -6}^{+   7}$  & $   2.5_{  -0.5}^{+   0.5}$  & $  1.72_{ -0.03}^{+  0.03}$  & $  10_{  -2}^{+   2}$  &  $ 478.3\,(475)$\\
$3$        &$ 4.5$ -- $ 6.3$ & $ 305_{ -78}^{+  88}$  & $  17_{  -4}^{+   6}$  & $  30_{  -3}^{+   3}$  & $   3.0_{  -0.3}^{+   0.3}$  & $  1.85_{ -0.04}^{+  0.05}$  & $  12_{  -3}^{+   3}$  &  $ 527.4\,(475)$\\
$4$        &$ 6.3$ -- $ 8.2$ & $ 424_{-104}^{+  94}$  & $  19_{  -5}^{+   6}$  & $  32_{  -2}^{+   2}$  & $   3.0_{  -0.3}^{+   0.3}$  & $  1.85_{ -0.04}^{+  0.04}$  & $  16_{  -3}^{+   3}$  &  $ 483.8\,(475)$\\
$5$        &$ 8.2$ -- $10.3$ & $ 142_{ -27}^{+  37}$  & $   7_{  -1}^{+   1}$  & $  25_{  -5}^{+   6}$  & $   1.6_{  -0.3}^{+   0.4}$  & $  1.88_{ -0.04}^{+  0.05}$  & $  17_{  -4}^{+   4}$  &  $ 515.8\,(475)$\\
$6$        &$10.3$ -- $13.2$ & $  86_{ -11}^{+  13}$  & $   5_{  -1}^{+   1}$  & $  16_{  -2}^{+   2}$  & $   1.0_{  -0.2}^{+   0.2}$  & $  1.88_{ -0.05}^{+  0.06}$  & $  10_{  -3}^{+   4}$  &  $ 471.1\,(475)$\\
$7$        &$13.2$ -- $16.3$ & $  82_{ -15}^{+  21}$  & $   4_{  -1}^{+   1}$  & $  18_{  -4}^{+   5}$  & $   1.0_{  -0.2}^{+   0.3}$  & $  1.92_{ -0.06}^{+  0.07}$  & $  13_{  -4}^{+   4}$  &  $ 534.1\,(475)$\\
$8^{(a)}$  &$16.3$ -- $20.6$ & $ 217_{ -53}^{+  70}$  & $   4_{  -1}^{+   2}$  & $  30_{  -3}^{+   4}$  & $   1.4_{  -0.2}^{+   0.2}$  & $  1.99_{ -0.06}^{+  0.08}$  & $  16_{  -3}^{+   5}$  &  $ 578.1\,(484)$\\
$9$        &$20.6$ -- $27.5$ & $ 592_{-397}^{+ 179}$  & $  10_{  -7}^{+   4}$  & $  26_{  -4}^{+   3}$  & $   1.0_{  -0.1}^{+   0.1}$  & $  1.97_{ -0.05}^{+  0.07}$  & $  11_{  -2}^{+   3}$  &  $ 604.2\,(475)$\\
$11^{(a)}$ &$37.1$ -- $53.2$ & $ 448_{ -229}^{+181}$  & $   5_{  -2}^{+   2}$  & $  33_{  -6}^{+   8}$  & $   0.4_{  -0.1}^{+   0.1}$  & $  1.98_{ -0.06}^{+   0.06}$ & $   5_{  -1}^{+   1}$  &  $ 751.1\,(477)$\\
$14$       &$82.7$ -- $88.1$ & $54_{-20}^{+40}$       & $ 0.3_{-0.2}^{+ 0.2}$  & $   6_{  -2}^{+   5}$  & $   0.06_{-0.04}^{+  0.04}$  & $  1.83_{ -0.11}^{+   0.13}$ & $ 1.1_{-0.1}^{+ 0.1}$  &  $ 1160.4\,(1436)$\\
\hline
\end{tabular}

Note: (a) Including LAT data
\label{t_2bbpl}
% \end{scriptsize}
\end{table*}

\begin{table*}\centering
\caption{Parameters of time resolved spectral fitting with BB + PL model in bin 10--14. Bin 12--14 have a 
simultaneous coverage with the XRT.}%\begin{scriptsize}
% \vspace{0.3in}
\begin{tabular}{ccccccccc}

\hline
Bin \# & Time interval (s) & $kT$\,(keV) &  $N_{\rm kT}$ & $\Gamma$  & $N_{\Gamma}$ &  PG-Stat (dof) &  PG-Stat(dof)$\rm _{PL}$  &  PG-Stat (dof)$\rm _{BBCPL}$ \\
\hline
\hline 
$10$ &$27.5$ -- $37.1$ & $   28_{ -7}^{+8 }$    & $   0.2_{  -0.1}^{+  0.1}$    & $ 1.52_{  -0.05}^{+   0.05}$    & $   2.3_{-0.5}^{+0.6}$  &  $ 638.1\,(469)$   &  $ 642.1\,(471)$   &  $ 640.1\,(468)$\\
$11$ &$37.1$ -- $53.2$ & $  85_{-19}^{+23 }$    & $   1.1_{  -0.2}^{+  0.2}$    & $ 1.98_{  -0.05}^{+   0.05}$    & $   5.6_{-1.0}^{+1.2}$  &  $ 769.3\,(479)$   &  $ 830.4\,(480)$   &  $ 757.1\,(478)$\\
$12$ &$53.2$ -- $67.7$ & $  13_{  -3}^{+ 4}$    & $   0.11_{  -0.04}^{+  0.04}$    & $ 1.41_{  -0.02}^{+   0.02}$    & $   0.75_{-0.03}^{+0.03}$  &  $ 1497.1\,(1438)$ &  $ 1505.4\,(1440)$ &  $ 1498.1\,(1437)$\\
$13$ &$67.7$ -- $82.6$ & $  11_{-4}^{+6}   $    & $ 0.05_{-0.04}^{+0.04}$       & $ 1.55_{  -0.03}^{+   0.03}$    & $   1.09_{-0.04}^{+0.04}$  &  $ 1571.1\,(1438)$ &  $ 1572.7\,(1440)$ &  $ 1565.7\,(1437)$\\
$14$ &$82.6$ -- $88.1$ & $  50_{-24}^{+49} $    & $ 0.2_{-0.1}^{+  0.2}$        & $ 1.73_{  -0.08}^{+   0.08}$    & $   1.04_{-0.08}^{+0.08}$  &  $ 1162.2\,(1438)$ &  $ 1164.8\,(1440)$ &  $ 1162.3\,(1437)$\\
\hline
\end{tabular}

\label{t_bbpl}
% \end{scriptsize}
\end{table*}

\label{lastpage}

\end{document}